\documentclass[twocolumn,aps,prx]{revtex4}

\usepackage{amsmath}    
\usepackage{physics}

\usepackage{amsfonts}
\usepackage{amssymb}
\usepackage[colorlinks=true,citecolor=blue,linkcolor=red]{hyperref}
\usepackage{graphicx}
\usepackage{bbold}					
\usepackage[dvipsnames]{xcolor}

\usepackage{ulem}
\usepackage{bm} 
\usepackage{subfigure}
\usepackage{dcolumn}
\usepackage{mathtools}

\usepackage{array,booktabs,ragged2e}	
\usepackage{soul}						
\usepackage{cancel}						
\usepackage{tabularx}
\usepackage{multirow}
\usepackage{hhline}
\usepackage{adjustbox}
\usepackage{sidecap}
\usepackage{threeparttable}

\usepackage[utf8]{inputenc}
\usepackage[T1]{fontenc}

\def\bf#1{\textbf{#1}}

\newcommand{\diff}{\mathrm{d}}

\newcolumntype{P}[1]{>{\raggedleft\arraybackslash}p{#1}}
\newcolumntype{R}[1]{>{\centering\arraybackslash}p{#1}}

\begin{document}

\title{Complex field, temperature and angle dependent Hall effects from intrinsic Fermi surface revealed by first-principles calculations}

\author{ShengNan Zhang$^{1}$}
\author{Zhihao Liu$^{1,2}$}
\author{Hanqi Pi$^{1,2}$}
\author{Zhong Fang$^{1,2,3}$}
\author{Hongming Weng$^{1,2,3}$}
\email{hmweng@iphy.ac.cn}
\author{QuanSheng Wu$^{1,2}$}
\email{quansheng.wu@iphy.ac.cn}

\affiliation{${^{1}}$Beijing National Laboratory for Condensed Matter Physics and Institute of physics,
Chinese Academy of Sciences, Beijing 100190, China}
\affiliation{${^{2}}$University of Chinese academy of sciences, Beijing 100049, China}
\affiliation{${^{3}}$Songshan Lake Materials Laboratory, Dongguan, Guangdong 523808, China}
\date{\today}

\begin{abstract}

The Hall effect, ever intriguing since its discovery, has spurred the exploration of its phenomena, intensified by advances in topology and novel materials. Differentiating the ordinary Hall effect from extraordinary properties like the anomalous Hall effect (AHE) is challenging, especially in materials with topological origins. In our study, we leverage semiclassical Boltzmann transport theory and first-principles calculations within the relaxation time approximation to analyze Hall effects comprehensively. We have found that the complex magnetic field dependence of ordinary Hall effect, including the sign reversals, appearing of plateau and non-linearity, can be understood and reproduced by our approach both for multi-band models and realistic topological materials of ZrSiS and PtTe$_2$. The Hall resistivity versus temperature and magnetic fields can be well scaled, similar to Kohler's rule for longitudinal resistivity. This methodology can also accurately models the angular dependent Hall effects such as  planar Hall effects of  bismuth. These findings indicate that the dependencies of various Hall and magnetoresistance on magnetic fields are mainly determined by the details of Fermi surface and the relaxation time. The intrinsic Fermi surface determines the carriers' density, type and velocity while the later is mostly influenced by extrinsic factors, such as quality of sample with defects, impurities and domains. This insight might simplify the understanding of several seemly complex transport phenomena in nonmagnetic materials with no need for hypotheses of other sophisticated mechanism, such as magnetization caused AHE, Lifshitz transition caused change in carrier type, exotic orders like charge density wave and some delicate scattering of carriers with chiral or nonreciprocal dependence. In order for the completeness of the discussion of Berry curvature, we also discussed the Hall effects when the Berry curvature is considered in the magnetic materials. 

\end{abstract} 

\maketitle

\section{Introduction}
The Hall effect, ever since its discovery by Edwin H. Hall in 1879~\cite{Hall1879}, has remained one of the most fundamental and notable phenomena, consistently inspiring new research. It is usually called as ordinary Hall effect since its counter part, the anomalous Hall effect (AHE) was discovered soon after, which showed the magnitudes ten times larger in ferromagnetic iron than in nonmagnetic conductors~\cite{Hall1881}. The AHE phenomenon remained elusive for a century, because it is deeply rooted in advanced concepts of topology and geometry that have been formulated only in recent times~\cite{AHERMP}. Consequently, topology has revitalized the field of transport, exhibiting a series of new phenomena, like quantum Hall effect~\cite{KlitzingQHE}, spin Hall effect~\cite{SHEJEPT, SHEPLA}, anomalous Hall effect~\cite{Hall1881, AHERMP} , planar Hall effect~\cite{PHEprb, PHEscience, PHEGaMnAs}, negative magnetoresistance~\cite{NMRnaturephy}, chiral anomaly~\cite{CApr, CANuovoCim} and many others, which are regarded as indicators of new physics. 

The AHE phenomenon was firstly noted in ferromagnetic materials\cite{Hall1881} but later was proposed in nonmagnetic materials~\cite{Maryenko2017,AHEZrTe5Liang} as well. Given that ordinary and anomalous Hall effect always occur together with similar shape of the resistivity curves and close order of magnitude in nonmagnetic materials, especially in high field limit, it is very difficult to distinguish them from each other. For anomalous Hall resistivity (conductivity), scientists usually examine its dependence on longitudinal resistivity (conductivity) to parse it~\cite{AHERMP}, since it could be classified as either intrinsic~\cite{Chang1996, Sundaram1999, Bohm2003, XiaoDiRMP} or extrinsic origin. However, for ordinary Hall effect there is rare study on its scaling behaviors. In other words, there are relatively clear ways to identify the AHE of ferromagnetic and topological materials. But it is a challenge to distinguish the resistivity curve with nonlinear and/or sign reversal features of non-ferromagnetic materials from AHE. To clarify this, understanding the Hall effect that caused by the Lorentz force and contributions from Berry curvature~\cite{AHEFeprb73, Schadjpcm1998, AHEFeprb09} becomes imperative, and further serves as a systematic method to study the Hall effect.

Another captivating phenomenon is the planar Hall effect (PHE). Characterized by an oscillating transverse voltage upon rotating the magnetic field in the plane determined by the current and Hall bar setup surface, the PHE has generated attention across both topologically trivial and nontrivial materials. Initially, the PHE was identified in ferromagnetic materials like $\rm GaMnAs$, $\rm Fe_3Si$, and $\rm Fe_3O_4$, attributed to anisotropic magnetoresistance resulting from significant spin-orbit coupling with magnetic order~\cite{PHEGaMnAs, PHEFe3Si2005, PHEFe3O42008}. However, recent studies have detected PHE in new-found topological insulators and Weyl semimetals, such as $\rm Na_3 Bi$\cite{PHENa3Bi}, $\rm ZrTe_5 $ \cite{PHEZrTe5}, GdPtBi\cite{PHEGdPtBi,PHEGdPtBiprb}, $\rm Bi_{2-x}Sb_xTe_3$\cite{PHEBiSbTe}, $\rm Cd_3As_2$~\cite{PHECd3As2}, $\rm ZrTe_{5-\delta}$ ~\cite{PHEZrTe5prb}, $\rm WTe_2$ \cite{PHEWTe2}, $\rm SmB_6$\cite{PHESmB6}, Te ~\cite{PHETe}, and $\rm Sr_3SnO$ ~\cite{PHESr3SnO}.  There are various proposed mechanisms behind PHE including anisotropic Fermi surfaces, time-reversal symmetry breaking in topological surface states, and the chiral anomaly in Weyl semimetals. Wang et al. further enriched this understanding by shedding light on the orbital contribution to PHE, pointing out that traditional theories might overlook this aspect~\cite{PHEyang2022}. Zhou and colleagues observed an anomalous in-plane Hall voltage persisting up to 380 K, which exhibits an unusual anisotropy magnetic field angular dependence. They demonstrate this in-plane Hall effect arises from an out-of-plane Berry curvature induced by the in-plane magnetic field~\cite{InplaneNature}. These fresh insights into the field of Hall effects suggest that there's still much to uncover and explore. Moreover, the advantages of PHE sensors include their thermal stability, high sensitivity, and very low detection limits, supporting a wide range of advanced applications such as nano-Tesla magnetometers, current sensing, and low magnetic moment detection. This makes them promising for applications in the integrated circuits industry.

The temperature dependence of resistivity is always an important and intriguing phenomenon to understand the magneto transport properties of materials. Kohler's rule~\cite{Kohlerrule, Kohlerfig}, a magneto resistivity scaling behavior that collapses a series of magnetoresistance curvers with different temperatures onto a single curve, has long been established and studied. Recently, researchers reported a transformative "turn-on" phenomenon within the resistivity-temperature curve $\rho (T)$, a unique feature of materials with extremely large magnetoresistance~\cite{MRWTe2nature, turnonKohler, turnonLasb, MRalphaWP2, turnonYBi, MRTaSe3, MRMoO2, turnonPdTe2, ExtendedKohler}. This phenomenon was once considered as a magnetic-field-driven metal-insulator transition, but perfectly reproduced and explained by Kohler's rule, as demonstrated by the work of Wang and colleagues~\cite{turnonKohler}. The successful application of Kohler's rule to longitudinal resistivity(or MR) raises an intriguing query: Does a scaling law exist for transverse resistivity, such as Hall resistivity? Further exploration and study in this fascinating area of research promise to yield new insights and deepen our understanding of the fundamental principles that govern magnetotransport behavior.

Despite the increasing interest in topological properties and its correlation with transport phenomena,  the Hall effect - rooted in semiclassical Boltzmann theory - remains a cornerstone of magnetotransport understanding. Our previous work~\cite{MRZhangprb} demonstrates that the Lorentz force and Fermi surface geometry play crucial roles in magnetoresistance. Combining semiclassical Boltzmann theory with first principles calculations has proven powerful in predicting the magnetoresistance of real materials, as observed from experiments. Examples include the non-saturating MR of bismuth, Cu~\cite{MRZhangprb}, SiP$_2$~\cite{MRSiP2}, $\alpha-$WP$_2$~\cite{MRalphaWP2}, MoO$_2$~\cite{MRMoO2}, TaSe$_3$~\cite{MRTaSe3}, ReO$_3$~\cite{MRReO3}, as well as large anisotropic MR of type-II Weyl semimetal WP$_2$~\cite{MRZhangprb}, ZrSiS~\cite{MRZrSiS}, the narrow-gap semiconductor ZrTe$_5$ ~\cite{pi2024_ZrTe5}, and magnetic materials like Co$_3$Sn$_2$S$_2$~\cite{zhliu2024}, among others. Yet, it is evident that comprehensive and systematic studies into Hall effect aspect are notably lacking, especially from first principles calculation studies. Addressing this gap in knowledge is crucial to fully comprehend the various manifestations of the Hall effect, whether in topologically trivial or nontrivial materials.

In this study, we present a comprehensive first principles exploration of the Hall effect, encompassing {\bf {field, temperature and angular dependence}, to complete the story of magnetoresistance within the context of Boltzmann transport theory and relaxation time approximation in our previous work~\cite{MRZhangprb}. The Hall effect is influenced by the geometry of the Fermi surface, temperature, and external magnetic fields, thus our discussion will also be concentrated in these three directions. 

First, we discuss the {\bf {field dependence}} of the Hall resistivity. The transport properties of materials are determined by carriers near the Fermi surface, and precise calculation requires a combination of first-principles method and the Boltzmann transport approach. However, a multi-band model can portray the physical picture in a sense, so we adopt several parameters (concentration and mobility of charge carriers) to qualitatively simulate the behavior of the Fermi surface. It is important to note that the multi-band model is merely an abstraction of the characteristics of the Hall effect, such as the sign reversal of the Hall resistivity, nonlinearity, among others, and does not represent the entire transport story. 
For some nonmagnetic materials, e.g., PtTe$_2$ and ZrSiS, their Hall resistivity show some distinct features similar to the AHE in ferromagnetic metals~\cite{AHEFeprb73, AHEFeprb09} and in the topological insulator $\rm ZrTe_5$~\cite{AHEZrTe5Liang}, such as the nonlinear slope and/or sign reversal Hall resistivity. Follow the multi-band models simulation, the shape of Hall resistivity curve is determined by the interplay between mobility and concentration of charge carriers. We then take a first-principles calculation on typical two-band material PtTe$_2$ to demonstrate these features, similar to the AHE, could be calculated and interpreted in the scope of the Boltzmann transport theory, resulting in surprisingly good agreement with experiment measurements.

Second, we discuss the {\bf {temperature dependence} of the Hall resistivity. Temperature dependence mainly comes from the Fermi distribution function and relaxation time. In semiconductors and some semimetals with fewer carriers, temperature can significantly affect carrier concentration and, thereby, transport properties; However,for metals with a high concentration of charge carriers, the impact of temperature on the Fermi surface can be negligible.  In this work, we only discuss the temperature dependence of Hall resistivity caused by the relaxation time. We successfully extend the Kohler's rule to the Hall effects to discuss the scaling behavior of the Hall effect with temperature, analogue to the Kohler's rule of magnetoresistance. Consequently, we name it as Kohler's rule of Hall resistivity. To demonstrate this, we chose PtTe$_2$ and ZrSiS as representative materials to elucidate the scaling behavior of their Hall resistivity relying on the magnetic field and temperature. Our theoretical calculations agree well with the corresponding experimental results, underscoring the veracity of this semiclassical approach. To our knowledge, Kohler's rule of Hall resistivity has never been made explicitly, but it is extremely important to help understanding the Hall resistivity scaling of multi type charge carrier materials.

Ultimately, we discuss the {\bf {angular dependence} of the Hall resistivity. When the direction of the magnetic field is within the plane of the sample, our method can also reproduce and explain the planar Hall effect in bismuth. In all, our study provides a methodology for systematically accounting transverse resistivity from semiclassical Boltzmann transport over a wide range, encompassing nonlinear properties, scaling behaviors, and Planar Hall resistivity originating from Fermi surface geometry. Moreover, the Chambers equation, as the spirit of the semiclassic magnetotransport theory, always lies at the heart of the calculation of various resistivity. 

Our paper is organized as follows. In Section II, we review our computational methodology. Section III considers a few multi-band models to specify the nonlinear features of Hall resistivity and apply it to real meterial PtTe$_2$. Section IV discusses the scaling behavior of Hall resistivity concerning temperature in PtTe$_2$ and ZrSiS materials. Section V focuses on the planar Hall effect. Section VI discusses how the Berry curvature affects the Hall resistivity curves in magnetic materials.  Finally, Section VII summarizes our work.

\section{METHODOLOGY}
In this work, we employ a sophisticated methodology combining first principles calculations and semiclassical Boltzmann transport theory to accurately compute magnetoresistance and Hall effects in models and real materials, ZrSiS, PtTe$_2$ and bismuth. First we construct tight-binding models using the first principles calculation software such as VASP~\cite{Vasp1, Vasp2} along with Wannier function techniques such as Wannier90~\cite{NMwannier90}. This approach allows us to obtain Fermi surfaces consistent with first principles calculation results, providing an accurate representation of the electronic structure of the materials under study.

After constructing tight-binding models, we apply the Chambers equation for rigorous treatment of the Lorentz force generated by the magnetic field acting on electrons within the Fermi surface. By precisely solving the semiclassical Boltzmann equation in the presence of a magnetic field, we get the conductivity for each energy band $\sigma_n(B)$ in the investigated materials. The resistivity is obtained by inverting the total conductivity tensor, which is derived by summing the conductivities across all energy bands $\sigma(B)=\sum_n \sigma_n(B)$. The emergence of Berry curvature leads to an anomalous Hall conductivity (AHC) $\sigma^A$, which is related to the magnetization, typically increasing with the magnetization. Therefore, in magnetic materials, we need to account for the contribution of this AHC to the total conductivity $\sigma(B)=\sum_n \sigma_n(B)+\sigma^A$ . In this work, we consider only the zeroth-order correction to conductivity due to Berry curvature, excluding the contribution of Berry curvature induced by the magnetic field. Our detailed computational methods are thoroughly explained in Reference~\cite{MRZhangprb, Liuyiprb} and implemented within the WannierTools software package~\cite{WUWT}.

Furthermore, we construct and analyze tight-binding toy models to explore the underlying physics governing magnetoresistance and Hall effects. The development and implementation of these toy models, as well as the results obtained, are extensively described in the Supplemental Material(\cite{supp}). This comprehensive methodology enables a deeper understanding of the complex interplay between electronic structure, magnetic fields, and transport properties in a broad of materials.

\section{Field dependence: nonlinearity of resistivity curves}

In the early stages of Hall effect research, it was generally accepted that the Hall resistivity $\rho_{yx}$ exhibits a linear relationship with the magnetic field $B_z$, in accordance with the Lorentz force acting on individual charge carriers. However, the situation becomes more complex when both electron and hole charge carriers are present, leading to a nonlinear Hall resistivity curve. This phenomenon persists even in the absence of a complex Fermi surface structure and has been documented in previous studies ~\cite{Fawcett1964,MRZhangprb}. Such nonlinearity in the Hall curve cannot always be captured by simplistic two-band or multi-band models when it comes to real materials. In these instances, the nonlinear Hall resistivity curves are often attributed to the anomalous Hall effect~\cite{AHEZrTe5Liang}, which is induced by the magnetic field. Regarding the similarity and perplexity features of the ordinary and anomalous Hall resistivity~\cite{AHEZrTe5Liang}, a comprehensive analysis is essential to clarify these ambiguity and draw a global picture. Therefore in this section, we construct several isotropic multi-band toy models to simulate Hall resistivity with typical behaviors, including nonlinearity, sign reversal, and plateaus. 

Although simple multi-band models can capture certain nonlinear characteristics of the Hall curve, the Fermi surfaces of real materials are often quite complex, featuring numerous characteristics of charge carriers that cannot be described by these simple models. At this point, our first-principles calculation methods become necessary. In this section, we also discuss the application of our method to the real material PtTe$_2$.

\subsection{Field dependence: a two-band model}
\begin{figure*}[!htpb]
    \centering
    \includegraphics[width=17cm]{ 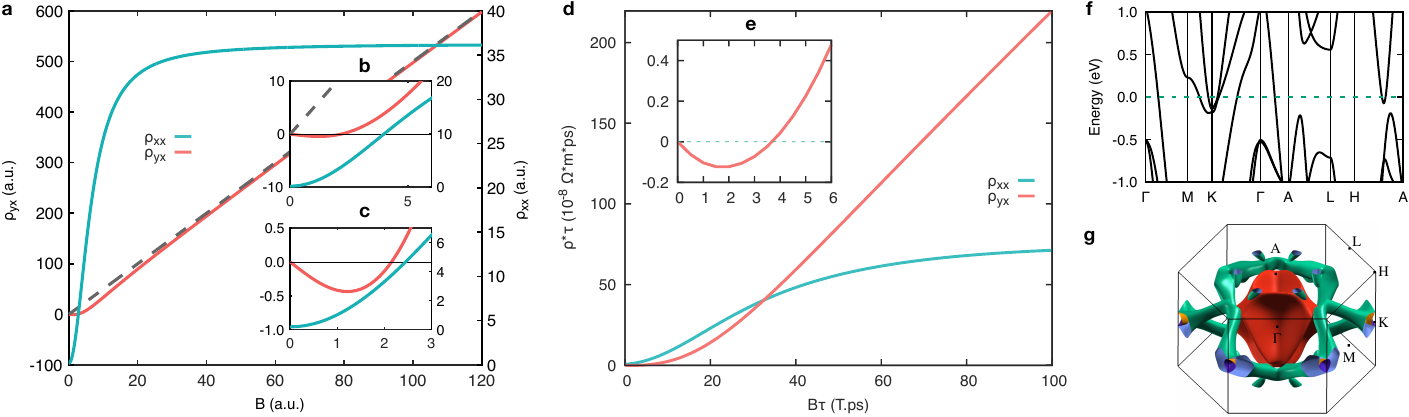}
    \caption{Longitudinal $\rho_{xx}$ (cyan line) and Hall resistivity $\rho_{yx}$ (red line) of  (a) a 2-band model and (d) a material PtTe$_2$. (a) Parameters for the 2-band model are $n_e=1, n_h=1.2, \mu_e=4, \mu_h=1$ satisfying the conditions $n_e<n_h, \mu_e>\mu_h, n_h\mu_h^2<n_e\mu_e^2$. The grey dashed lines visible in (a) represent the Hall resistivity in the scenario of a high-limit magnetic field. Insets (b) (c) and (e) are enlarged views of the main panel, each within a different magnetic field range. Please note for (a)-(c), the left and right vertical axis are corresponding to $\rho_{yx}$ and $\rho_{xx}$ respectively. (f) (g) the energy band structures and Fermi surfaces of PtTe$_2$ respectively.}
    \label{fig1}
\end{figure*}

We begin with the isotropic two-band model to mimic a system with an electron pocket and one hole pocket, which involves four parameters: $n_e$ and $n_h$ representing the densities of electron and hole charge carriers, and $\mu_e$ and $\mu_h$ denoting the mobilities of electrons and holes, respectively.
The longitudinal and transverse resistivities for this model are given by the following well-known expressions:
\small
 \begin{align}
    &\rho_{xx}=\frac{1}{e} \frac{(n_e\mu_e+n_h\mu_h)+(n_e\mu_h+n_h\mu_e)\mu_e\mu_h B^2}{(n_e\mu_e+n_h\mu_h)^2+(n_h-n_e)^2\mu_e^2
    \mu_h^2B^2} \nonumber\\
    &\rho_{yx}=\frac{B}{e} \frac{(n_h\mu_h^2-n_e\mu_e^2)+(n_h-n_e)\mu_e^2\mu_h^2 B^2}{(n_e\mu_e+n_h\mu_h)^2+
    (n_h-n_e)^2\mu_e^2\mu_h^2B^2}\label{twobandeq} 
\end{align}
\normalsize
where we take the convention such that the positive value of $\rho_{yx}(B\rightarrow \infty)$ indicates net hole concentration as $n_h > n_e$. 

Despite its apparent simplicity, this model is able to reproduce results that closely agree with many experiments. It is frequently employed to fit MR and Hall resistivity in experiments, enabling the determination of charge carrier concentration and mobility of real materials. It is noteworthy that identical results can be achieved if one utilizes first principles calculations combined with the Boltzmann transport method to simulate isotropic multi-band model ~\cite{supp}.

The competition between constant and magnetic field dependent terms in Eq.~(\ref{twobandeq}) suggest that $\rho_{xx}$ and $ \rho_{yx}$ may exhibit distinct behaviors under extremely low and high magnetic fields. Meanwhile, the interplay between different charge carriers, i.e., concentration and mobility, would change the concrete form of the Hall resistivity curves.


At low magnetic fields ($B \rightarrow 0$),  the longitudinal resistivity $\rho_{xx} (B \rightarrow 0) \propto \frac{\mu_e \mu_h B^2}{e(n_e\mu_e+n_h\mu_h)}$, still exhibiting a parabolic dependence on the magnetic field. The Hall resistivity $\rho_{xy}$ is given by $ \frac{(n_h\mu_h^2-n_e\mu_e^2)}{e(n_e\mu_e+n_h\mu_h)^2} B$, which is proportional to $B$, and the sign of slope of the Hall curve depends on $(n_h\mu_h^2-n_e\mu_e^2)$. At very high magnetic fields ($B \rightarrow \infty$), $\rho_{xx}$ saturates, and the Hall resistivity still scales linearly as $\rho_{xy} \propto \frac{B}{e(n_h-n_e)}$, of which the sign of slope depends on $(n_h-n_e)$. This is distinct from the low-field case. Accordingly, the Hall resistivity curve may possess completely opposite slopes in the low and high magnetic field limits, generating various curve shapes. For convenience, we summarize the results of our two-band model in Table A1 in Supplementary materials~\cite{supp}.

Next, we shall present two examples to gain specific insights into various features of Hall resistivity originating from the interplay between different charge carriers. Consider two types of charge carriers with the following concentration and mobility values: $n_e = 1$, $\mu_e = 4$, $n_h = 1.2$, and $\mu_h = 1$. In this case, we have $n_e < n_h$, $\mu_e > \mu_h$, and $n_e\mu_e^2 > n_h\mu_h^2$, which is expected to result in distinct Hall resistivity slopes under low and high magnetic field. We plot the longitudinal and Hall resistivity $\rho_{xx}, \rho_{yx}$ for this case in Fig.\ref{fig1}. To fully understand their behaviors, it is essential to discuss them on different magnetic field ranges, namely low, moderate, and high magnetic fields. When we consider a large magnetic field scale (up to 120), we observe that the Hall resistivity(red line) shows a linear dependence on the magnetic field, while the longitudinal resistivity(blue line) saturates quickly(Fig.\ref{fig1}(a)). This is a common observation for semimetals or semiconductors, precisely in line with our intuition. However, this is not the integrity physical picture. When enlarging the magnetic field axis, we discover entirely different behaviors in the low and moderate magnetic field regime. This is plotted in the inset of Fig.\ref{fig1}(b) and (c).

Within the low magnetic field ranging from $0$ to $3$ shown as Fig.\ref{fig1}(c), the longitudinal resistivity $\rho_{xx}$(blue line) exhibits a nearly (or sub) parabolic scaling, while the Hall resistivity $\rho_{xy}$(red line) shows a sign-reversal feature, as mentioned previously. The slope of the Hall resistivity initially appears positive due to $(n_h\mu_h^2-n_e\mu_e^2) > 0$ at low magnetic field, and then changes to negative as the magnetic field increases, determined by $n_h-n_e < 0$. When observing the resistivity curve at a moderate magnetic field scale, e.g., $0$ to $5$, as depicted in Fig.\ref{fig1}(b), the sign-reversal feature becomes less distinct but appears similar to the anomalous Hall effect (indicated by the black dashed line and the area between it and the red curve). This moderate magnetic field range is the most frequently used in experiments measurement.

\subsection{Field dependence: a real material PtTe$_2$}
To apply these analysis to real two-band materials, we choose a representative of type-II Dirac semimetal PtTe$_2$. Pavlosiuk and Kaczorowski~\cite{Ptte2nasci} have performed magnetic transport measurement on PtTe$_2$ and reported that the longitudinal resistivity depends on magnetic field of the power law $\rho_{xx} \propto B^{1.69}$, and that the Hall effect data exhibits a multi-band character with moderate charge carrier compensation. Both of these behaviors indicate its charge carriers deviate from perfect compensation. We reproduce these feature in our calculations, and plot both longitudinal and Hall resistivity curves in Fig.~\ref{fig1}(d). The band structure of PtTe$_2$, shown in Fig.~\ref{fig1}(f), exhibits complex structure near the Fermi energy and hence intricate Fermi surface displayed in Fig.~\ref{fig1}(g). This is consist with the quantum oscillation results in Ref.~\cite{Ptte2nasci}, which confirm more than three electron and hole pockets but only one electron and one hole bands dominating the transport near Fermi level. 

Now we shall compare our calculated results with the experimental measurements reported in Ref~\cite{Ptte2nasci}. The measured MR exhibits an sub-quadratic pow law dependence on magnetic field and remains unsaturated until $B=8$ T(Fig.2a in Ref~\cite{Ptte2nasci} ). In our calculations, the blue line of longitudinal resistivity $\rho_{xx}$ in Fig.~\ref{fig1}(d) also shows sub-quadratic dependence on magnetic field under $B\tau$ no more than 10 T$\cdot$ps but becomes saturated at very large field over $B\tau=50$ T$\cdot$ps. On the other hand, the Hall resistivity $\rho_{yx}$ exhibits a complex manner: at low temperature $T=2 \sim 25$ K as shown in Fig.~\ref{fig3} (c) (or Fig.6a in Ref~\cite{Ptte2nasci}), i.e., it starts as negative, drop to a minimum, then switches to positive, and continues to rise with increasing magnetic field. Our calculated results perfectly reproduce the sign reversal feature, as shown in Fig.~\ref{fig1}(e): the negative minimum of Hall resistivity appears around $B \tau=2$ T$\cdot$ps, and then changes to positive around $B \tau=4$ T$\cdot$ps, which agrees quite well with experimental measurements.

\begin{figure*}
    \centering
    \includegraphics[width=15cm]{ 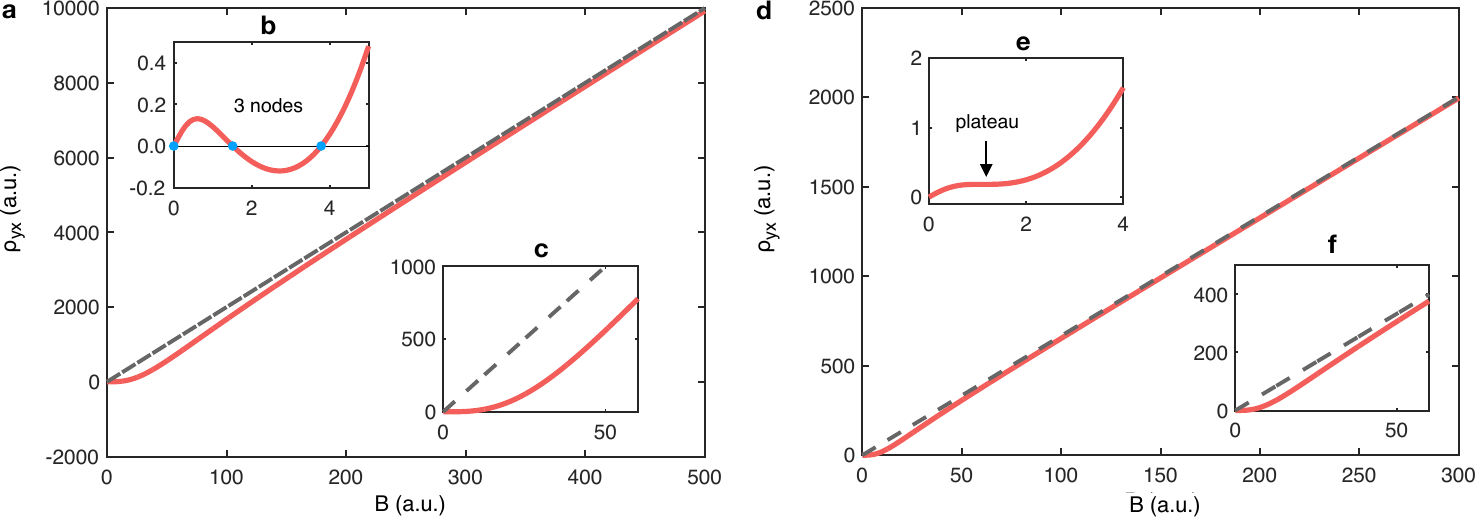} 
    \caption{Hall resistivity $\rho_{yx}$ of two 3-band models with 
    parameters $n_e=1$, $n_{h1}=0.5$, $n_{h2}=0.55$, $\mu_e=1$, $\mu_{h1}=0.5$, $\mu_{h2}=2.5$, $n_{h1}+n_{h2}>n_e$  for (a), exhibiting a 3-node structure, and  $n_e=1$, $n_{h1}=0.5$, $n_{h2}=0.55$, $\mu_e=1$, $\mu_{h1}=0.65$, $\mu_{h2}=2.5$  for (d), exhibiting a plateau structure. Insets (b) (c) and (e) (f) are enlarged views of the main panel. The grey dashed lines represent the Hall resistivity in the scenario of a high-limit magnetic field.}
    \label{fig2}

\end{figure*}

\subsection{Field dependence: Three-band model}

Furthermore, we go deeper to a three-band model, for example one type of electron charge carrier and two types of hole charge carriers. By explicitly writing their resistivity equations at low magnetic field ($B \rightarrow 0$) explicitly (see the detailed equations A.16-A.17 in ~\cite{supp}), we can see that the sign of the Hall resistivity is determined by the term $ -n_e\mu_e^2+n_{h1}\mu_{h1}^2+n_{h2}\mu_{h2}^2$, owing to the interplay of both concentration and mobility of multiple charge carriers. Regarding the high magnetic fields case ($B \rightarrow \infty$), the Hall resistivity is written as, 

\begin{align}
    &\rho_{yx}=\frac{B }{e(-n_e+n_{h1}+n_{h2})}
    \label{threebandrho}
\end{align}
which is evident that the sign of Hall resistivity is only determined by the net concentration of multiple charge carriers $-n_e+n_{h1}+n_{h2}$, implying that the net charge quantity dictates the slope of Hall resistivity under high magnetic fields. Based on the distinct form of Hall resistivity under low (see equation A.16-A.17 in ~\cite{supp}) and high magnetic fields (Eq.\ref{threebandrho}), it is natural to conclude that the Hall resistivity may change sign when altering the magnitude of magnetic field. 

The physical origin of this sign reversal is the competition between different type of charge carriers. We could understand it qualitatively as follows. Through the previous analysis, the Hall resistivity finally shows the expected sign at high field limit, i.e., the net charge concentration. However, at low magnetic field it sensitively depends on practical motion trajectory. We take the standard definition of a dimensionless quantity $\omega \tau = \frac{e}{m} B \tau$ to mark achievement of the cyclotron motion, from which one finds that charge carriers with heavier masses will complete one circle more slowly than the lighter one. Consequently, the lighter charge carrier exhibits their charge onto the sign of Hall curve earlier than the heavier ones. As all charge carriers have completed numerous circles under high magnetic field, the net quantity of charge carriers takes over to determine the sign of Hall resistivity. If the Hall resistivity signs differ in low and high magnetic field scenarios, then a sign reversal must occur.

We shall present two examples to illustrate how the Hall resistivity sensitively depends on the details of the charge carriers, as shown in Fig.~\ref{fig2}. From the three-band model, one can easily determine the slope of the Hall resistivity at very low and high magnetic fields, but not in between. Here, we consider two cases that both satisfy $-n_e\mu_e^2+n_{h1}\mu_{h1}^2+n_{h2}\mu_{h2}^2 > 0$ and $-n_e+n_{h1}+n_{h2} > 0$. As shown in Fig.~\ref{fig2}(a) and (d), the Hall resistivity is plotted with a large magnetic field scale view, and both cases exhibit roughly positive Hall resistivity. First, let's consider the concentration and mobility of charge carriers with $n_e = 1$, $\mu_e = 1$, $n_{h1} = 0.5$, $\mu_{h1} = 1$, and $n_{h2} = 0.55$, $\mu_{h2} = 2.5$. In Fig.\ref{fig2}(b), the Hall resistivity initially shows a positive slope, but soon changes to negative and then back to positive again, resulting in sign reversal and 3-node oscillation features at low magnetic field regime. However, if we plot the Hall resistivity within the intermediate magnetic field range, a feature resembling the anomalous Hall effect occurs around the origin, as shown in Fig.~\ref{fig2}(c). As a comparison case shown in Fig.~\ref{fig2}(d), we only modify the mobility of one type of hole charge carriers from $\mu_{h1} = 0.5$ to $\mu_{h1} = 0.65$, while keeping other parameter unchanged. This still satisfies the condition that the slopes of Hall resistivity is positive under very low and high magnetic fields. However, instead of a 3-node oscillation feature, a plateau appears in the Hall resistivity, as shown in Fig.~\ref{fig2}(e). This can be understood by calculating the derivative of the Hall resistivity $\frac{\rm d \rho_{yx}}{\rm dB}$. Whether $\frac{\rm d \rho_{yx}}{\rm dB} = 0$ occurs once or twice determines the presence of a plateau or 3-node oscillation features, respectively. In all, we observe that ordinary Hall resistivity curve can appear features similar to the AHE ones, i.e., nonlinear slope and sign reversal, only considering semiclassic Boltzmann transport approach with multi types of charge carriers under intermediate magnetic field regime.

\section{Temperature  dependence: The Hall resistivity scaling behavior and temperature effect}

Kohler's rule is a particularly common phenomenon when one studies the scaling behaviors of MR, which states that the MR should be a function of the combined variable $\frac{B}{\rho_0}$, expressed as $\text{MR} = \frac{\Delta \rho}{\rho_0} = \text{F}\left(\frac{B}{\rho_0}\right)$. In principle, assuming that the scattering mechanisms of the material remain unchanged, the conductivity is expected to be proportional to the drift length divided by the relaxation time. As the magnetic field intensifies, the drift length correspondingly decreases, primarily because it is inversely proportional to the magnetic field.~\cite{Pippard}. Therefore, the conduction behavior of charge carriers can be scaled down when increasing the magnetic field, i.e., that all MR curves should collapse onto a single curve. This physical scaling picture is consistent with Kohler's rule.


\begin{figure*}
    \centering\includegraphics[width=17.5cm]{ 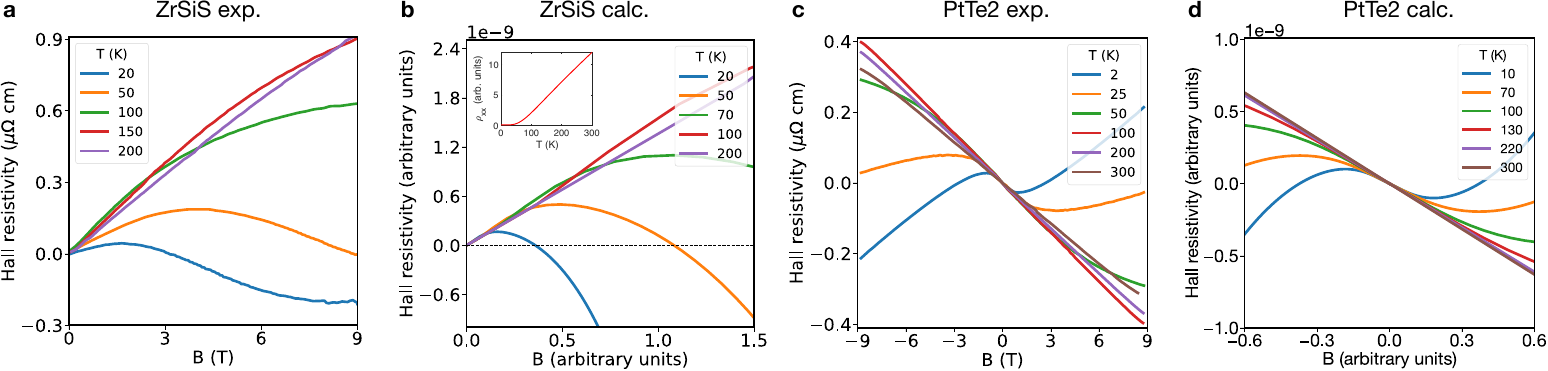}
    \caption{Field dependence of the Hall resistivity of ZrSiS and PtTe$_2$, (a) (c) Measured Hall resistivity at different temperatures for ZrSiS and PtTe$_2$, replotted from Ref.~\cite{singha2017} and Ref.~\cite{Ptte2nasci} respectively, (b), (d) Interpretations based on Kohler's rule, with Hall resistivity numerically calculated at low temperatures for both materials. Inset in (b): Longitudinal resistivity of ZrSiS at B=0 using the Bloch-Grüneisen model.}
    \label{fig3}
\end{figure*}

It is not difficult to find that the Chambers equation already embraces Kohler's rule after close and careful study. Chambers equation, represented by Eq. A.1 (~\cite{supp}), states that the product of resistivity and relaxation time is a function of the combined variable of magnetic field and relaxation time, i.e., $\rho \tau = f(B\tau)$. Recall that the zero-field resistivity is inversely proportional to the relaxation time as $\rho_0 \propto \frac{1}{\tau}$ in the Drude model, which is taken as a basic hypothesis in this section. Thus we have $B\tau \sim \frac{B}{\rho_0}$, and the Chambers equation recovers Kohler's rule. Moreover, the Chambers equation provides a more comprehensive framework for calculating all the elements of resistivity tensor, including the transverse resistivity elements; while Kohler's rule only concerns longitudinal resistivity. Namely, Chambers equation offers a more extensive formulation to investigate the scaling behavior of resistivity tensor. 

When the MR curves are scaled according to  different temperatures, they collapse onto a single curve for most materials. This implies temperature is another scaling variables besides magnetic field. The temperature does not exist explicitly in the resistivity expression but hidden in the zero field resistivity $\rho_0$ or relaxation time $\tau$. It's important to note that temperature impacts not only the relaxation time but also causes a shift in the Fermi energy and modifies the Fermi distribution function, leading to a comprehensive alteration of the transport properties. However, we are concerned with the effect of temperature on relaxation time in this work.

To better understand the scaling behavior of Hall resistivity with temperature, we briefly review the influence of thermal effects on resistivity through the relaxation time. Within the semiclassical transport framework, the relaxation time encompasses all scattering processes and is a key determinant of resistivity, which can be expressed as $\rho_0 \propto \frac{1}{\tau}$ in the absence of a magnetic field. Different scattering mechanisms lead to various relationships between resistivity and temperature. At very low temperatures, resistivity becomes temperature-independent ($\rho_0 \propto T^0$) due to solid impurity scatterings, which are unaffected by temperature. With increasing temperature, electron-electron (e-e) and electron-phonon (e-h) scatterings begin to dominate.

Typically, e-e scattering contributes to resistivity following $\rho_0 \propto \frac{1}{\tau} \propto T^2$. Given that the relaxation time $\tau$, originating from e-e scattering in metals at room temperature, is of the order of $10^{-10}$ seconds—approximately $10^4$ times larger than that of other scattering mechanisms—its impact on resistivity is relatively minor compared to others. E-h scattering also plays a crucial role in degrading currents, but its temperature dependence is more intricate. At relatively low temperatures (below the Debye temperature $\Theta_D$), the resistivity follows a power law relationship with temperature as $\rho_0 \propto T^5$. In contrast, at higher temperatures ($T > \Theta_D$), the resistivity demonstrates a linear temperature dependence, specifically $\rho_0 \propto T$.

Indeed, empirical descriptions provide qualitative insights, but for more precise quantitative analysis, we introduce the Bloch-Grüneisen (BG) model~\cite{ziman} in order to simulate the resistivity of a metallic system at zero magnetic field, which is written as, 
\small
\begin{eqnarray}
\rho_0 \equiv \rho(T, B=0)= \rho(T=0,B=0) \nonumber\\
+A(\frac{T}{\Theta_D})^n\int_0^{\Theta_D/T} \frac{t^n}{(e^t-1)(1-e^{-t})}dt \label{rho_BG}
\end{eqnarray}
\normalsize
where, $\rho(T=0,B=0)$ denotes the residual resistivity at zero magnetic field, and $\Theta_D$ is the Debye temperature. This BG equation effectively characterizes phonon scattering across the full temperature spectrum. In this context, the zero-field resistivity $\rho(T,B=0)$ can be approximated by a power law, $T^n$, in the low temperature regime (when $T \ll \Theta_D$), and transitions to a linear, $\sim T$, dependence in the high-temperature region. Utilizing the BG model allows us to quantitatively calculate the longitudinal resistivity as a function of temperature, considering the realistic scattering processes of materials.

To derive temperature dependent Hall resistivity curves varying with magnetic field from our calculated results $\rho \tau = f(B\tau)$, as per the Chambers equation, it is necessary to replace the relaxation time $\tau$ in the combined variables $B\tau$ and $\rho \tau$ with the temperature-dependent zero-field resistivity $\rho_0$. This approach is based on the hypothesis that $\tau \propto \frac{1}{\rho_0}$, essentially reversing the scaling analogy of Kohler's rule. The temperature-dependent zero-field resistivity $\rho_0$ can be computed using the Bloch-Grüneisen (BG) model as shown in Eq.~\ref{rho_BG} and is depicted in the inset of Fig.\ref{fig3}. For simplicity, we assume that each temperature corresponds to a constant relaxation time, ensuring that the scattering mechanism remains unchanged within the relaxation time approximation. Consequently, we compare our calculated Hall resistivity (as illustrated in Fig.\ref{fig3}(b)) with the experimental results (shown in Fig.\ref{fig3}(a)) for ZrSiS at various temperatures. The comparison reveals a relatively good agreement, capturing most of the essential features. 

For example, the Hall resistivity curves display distinct shapes at different temperatures, rather than a uniform pattern. Notably, at relatively low temperatures (up to $T = 50 \text{K}$), these curves exhibit a transition from positive to negative values under small magnetic fields. This transition is marked by an intercept on the $B$-axis, which indicates a change in the slope of the Hall resistivity. As the temperature increases, the bending of the curves becomes less pronounced. The behavior transitions from a downward trend to a relatively flat response and then eventually to a linear increase with the rising magnetic field.

In order to further clarify this point, we consider the calculated curves at very low temperatures, such as $T = 20 \text{K}$, where the relaxation time is large due to minimal scatterings. As the temperature rises, the relaxation time shortens, resulting in increase of intercept of the Hall resistivity curve on the $B$-axis due to the horizontal axis transforming from ($B\tau$) to $B$. This increasing intercept indicates that the positive portion of the Hall resistivity curve gradually becomes more prominent across the entire $B$-axis range. With the relaxation time significantly decreasing, by factors ranging from one to several dozen, the negative segment of the Hall resistivity curve is expected to progressively diminish at relatively high temperatures, as demonstrated in Fig.\ref{fig3}(b). From this observation, we infer that the series of Hall resistivity curves at different temperatures for ZrSiS are a manifestation of the relaxation time variation with temperature.

Expanding our analysis to additional materials, we compare the experimentally measured Hall resistivity of PtTe$_2$ with our corresponding calculated results in Fig.\ref{fig3}(c) and (d), demonstrating good agreement. By applying the same analytical approach used for ZrSiS, the behavior of Hall resistivity in PtTe$_2$ as temperature increases becomes readily understandable. Differing from the presentation in Fig.\ref{fig3}(a) and (b), in Fig.\ref{fig3}(c) and (d), we plot the Hall resistivity against the magnetic field from negative to positive direction, facilitating a direct comparison between our calculations (Fig.\ref{fig3}(d)) and experimental data (Fig.\ref{fig3}(c)). From lower temperatures ($T = 2 \text{K} \sim 25 \text{K}$) to higher ones ($T = 50 \text{K} \sim 300 \text{K}$), the interception of the Hall resistivity curves on the $B$-axis shows a progressive change from minimal to significant, eventually reaching infinity. We do not further detail the characteristic Hall curves of PtTe$_2$, as they originate from the same mechanisms observed in ZrSiS, and both sets of data show reasonable agreement with experimental results.

To our knowledge, the scaling behavior of Hall resistivity has been hardly studied both experimentally and theoretically. The reasons may be as follows; the scaling behavior of Hall resistivity is not that straight forward like the Kohler's rule for longitudinal resistivity. On one hand, the Hall resistivity linearly depends on magnetic field of material with single charge carriers for many metals, semimetals and semiconductors. Moreover, the behavior of the Hall resistivity under low and high magnetic field limit remains unaffected by temperature, as shown in Table A1~\cite{supp}, eliminating the need for scaling. On the other hand, in materials with multiple types of charge carriers that compensate each other, such as ZrSiS and PtTe$_2$, the Hall resistivity exhibits a complex function of magnetic field and temperature. This complexity often leads to nonlinear and sign reversal features, which are sometimes misinterpreted as Anomalous Hall Effect (AHE) or indicative of new physics. In contrast, longitudinal MR often shows a distinct power law dependence on magnetic field, making the scaling behavior more apparent and easier to summarize.

Additionally, we want to point out that our calculated Hall resistivity deviates from the experimental measurements at low temperatures (such as $T= 20, 50$K) and high magnetic fields in Fig.\ref{fig3}. This deviation suggests that the relaxation time varies not only with temperature but also with other parameters, such as the band index and the momentum $k$.

\section{Angular dependence: Planar Hall effects}

The Planar Hall effect (PHE), initially observed in ferromagnetic materials, has a magnitude of only a few percent, resulting from the anisotropic magnetoresistance (AMR) induced by spin-orbit coupling. Subsequently, a giant magnitude PHE was found in devices made of $\rm (Ga,Mn)As$ ~\cite{PHEGaMnAs}, with four orders of magnitude larger than that observed in ferromagnetic metals. As interests in topological materials grow, PHE has been detected in a variety of newly identified topological insulators and Weyl semimetals. Theoretical studies~\cite{ChiralAprb, ChiralAprl, PHEsysm} suggest a novel mechanism for PHE involving chiral anomaly and nontrivial Berry curvature, which are known to generate negative MR and in turn reinforce the PHE due to their enhancement of the AMR. 

A common feature of materials that exhibit the PHE, especially in ferromagnetic materials, is the presence of a pronounced AMR. Consequently, planar Hall resistivity has been understood from the perspective of AMR. In this context, the electric field components $(E_x, E_y)$ and the current density $J$ within the material's transport plane can be expressed as follows~\cite{AMR2008},
\begin{align}
&E_x = J \rho_{\perp} + J (\rho_{\parallel} - \rho_{\perp})cos^2 \theta \\
&E_y = J (\rho_{\parallel} - \rho_{\perp}) sin \theta cos \theta  
\end{align}
where $\rho_{\parallel}$ and $\rho_{\perp}$ represent the resistivity when the current is parallel and perpendicular, respectively, to the magnetic field within the film plane. In other words the longitudinal resistivity $\rho_{11}$ and the planar Hall resistivity $\rho_{12}$ can be explicitly expressed as,
\begin{align}
&\rho_{11} = \rho_{\perp} -  ( \rho_{\perp} -\rho_{\parallel})cos^2 \theta \label{rhoxx} \\
&\rho_{12} = - ( \rho_{\perp} -\rho_{\parallel}) sin \theta cos \theta \label{rhoph} 
\end{align}
where Eq.~\ref{rhoxx} describes the AMR, while Eq.~\ref{rhoph}  represents the planar Hall resistivity. To compare with the experimental results, we adopt the convention index for the resistivity elements, such that the $x-y$ coordinate system is fixed with the sample($\rho_{11}, \rho_{12}$ refer to $\rho_{xx}, \rho_{xy}$) and that the $x'-y'$ coordinate system is rotated with magnetic field(the $x'$ axis is parallel to the direction of magnetic field as $\rho_{\parallel}, \rho_{\perp}$ refer to $\rho_{x'x'}, \rho_{y'y'}$ respectively), which is shown in Fig.\ref{fig4} (j).

\begin{figure*}[!htpb]
\centering
    \includegraphics[width=17cm]{ 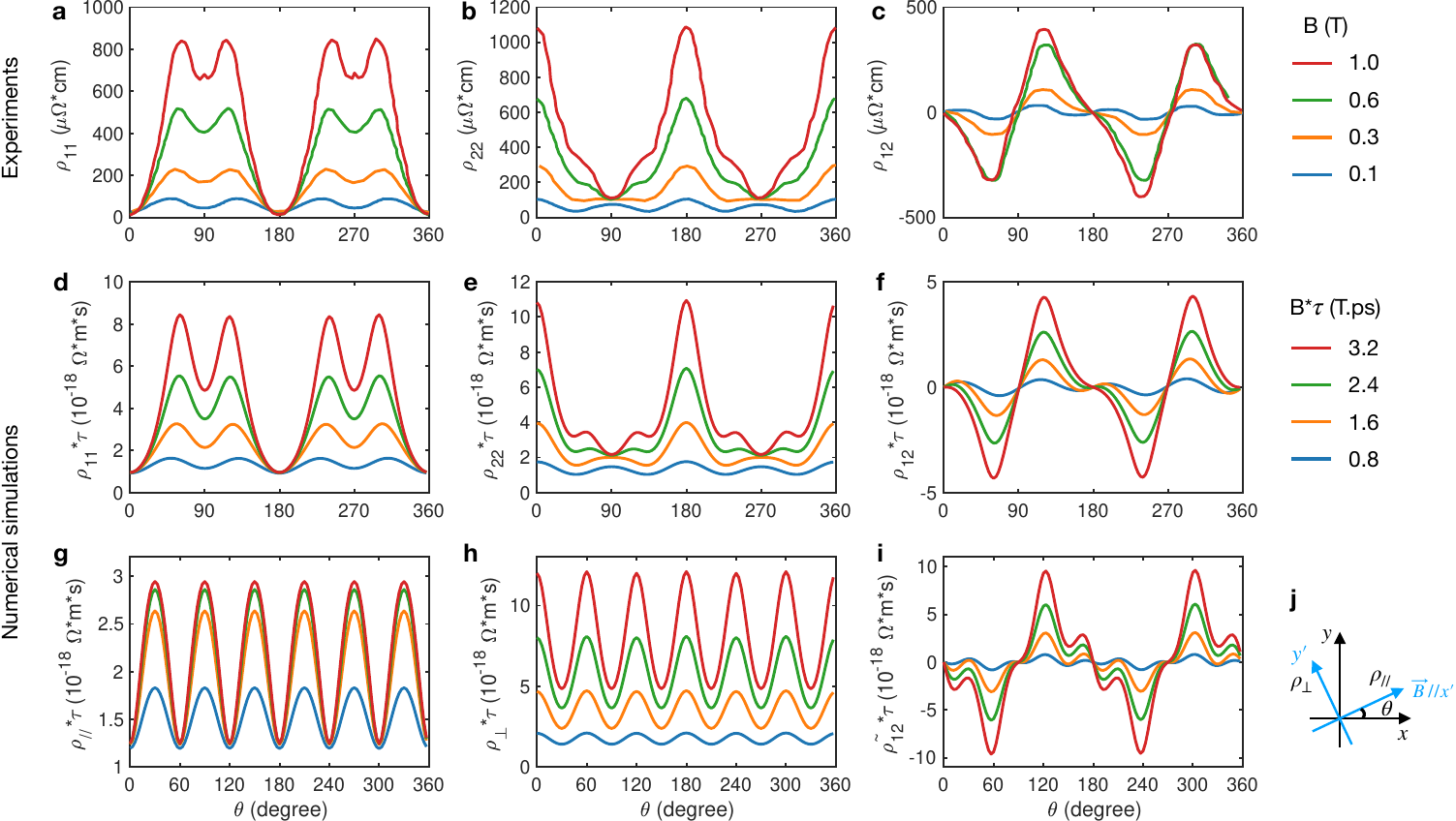}
    \caption{Experimental and numerically simulated AMR and PHE curves of bismuth. (a-c) Experimentally observed angular dependence of (a) $\rho_{11}$, (b) $\rho_{22}$ and (c) $\rho_{12}$ (PHE) with the current along the binary axis, reploted from Ref.~\cite{PHEBiprr}. (d-f) Numerical simulations of AMR and PHE corresponding to (a-c). (g-h) Representations of AMR where (g) $\rho_{\perp}$ and (h) $\rho_{\parallel}$ represent the resistivity with the magnetic ﬁeld perpendicular (90°) and parallel (0°) to the electric current. (i) Numerically simulated PHE calculated  based on (g) and (h) (see Eq.~\ref{rhoxx} and ~\ref{rhoph}), {excluding the off-diagonal contribution from the z-direction.} (j) Schematic diagram of the rotation transformation coordinate.}
    \label{fig4}
    
\end{figure*}

Meanwhile, PHE has also been observed in ordinary metals, such as bismuth. In 2020, a research group reported a significant PHE in bismuth, exceeding several $\rm m \Omega\cdot cm$~\cite{PHEBiprr}. They utilized a semiclassical model with an adjustable tensor to simulate the multivalley and anisotropic Fermi surface of bismuth. This model effectively described the PHE behavior before reaching the quantum limit, where the magnetic field strength is not extremely high. Based on this work, another group confirmed the PHE in bismuth~\cite{PHEBiprb} and expanded the semiclassical model to include the quantum limit one by incorporating the charge carrier concentration approach to Landau quantization.


Contrasting with their manual adjustable model, we employ a tight-binding Hamiltonian derived from density functional theory calculations, without relying on any adjustable parameters. Remarkably, this approach enables us to reproduce all the distinctive features of the PHE of bismuth, with the exception of the quantum limit. Our calculations accurately depict the resistivity behavior across a range of moderate magnetic fields and temperatures. Fig.\ref{fig4}(a)-(c) shows the experimental measurements, while Fig.\ref{fig4}(d)-(f) displays our calculated resistivity results \footnote{Here the resistivity elements $\rho_{11},\rho_{22},\rho_{12}$, shown in Fig.4(d)-(f),are calculated from Eq.10 rather than Eq.8 and Eq.9.}, demonstrating excellent agreement.

In Fig.\ref{fig4}(d), the longitudinal resistivity $\rho_{11}$ exhibits a periodic behavior, which is slightly deformed as magnetic field increases. For instance, at a very low magnetic field of  $B = 0.1 \ \text{T}$, $\rho_{11}$ shows four peaks of equal height and slight variations in the troughs, resembling a $\frac{\pi}{2}$ period, although the exact period is $\pi$. As increasing magnetic field, these four peaks at low fields transition into two m-shaped humps with central troughs. Additionally, in $\rho_{22}$ (or $\rho_{12}$), the peaks observed at $\theta = \frac{\pi}{2}, \frac{3\pi}{2}$ (or $\theta = \frac{\pi}{6}, \frac{7\pi}{6}$) at low fields evolve into valleys(part of valleys) with increasing magnetic field.

Similarly, the longitudinal resistivity $\rho_{22}$ and the planar Hall resistivity $\rho_{12}$ shown in Fig.\ref{fig4} (e) and (f) respectively, exhibit similar periodic tendencies as $\rho_{11}$. For the experimentally measured (Fig.\ref{fig4}(c)) and our calculated (Fig.\ref{fig4}(f)) planar Hall resistivity, at a low magnetic field of $B \tau=0.8$ T$\cdot$ps, the $\rho_{12}$ curve (in blue) displays an approximate $\frac{\pi}{2}$ period with distinct peaks and troughs. As the magnetic field increases, these peaks (or troughs) gradually flatten and disappear, eventually retaining only the troughs (or peaks) over a $\frac{\pi}{2}$ range. This trend is clearly observable from $B\tau = 1.6  \text{T} \cdot \text{ps}$ to $B\tau = 3.2 \text{T} \cdot \text{ps}$, corresponding to orange, green and red curves in Fig.\ref{fig4} (c) and (f). 

The intricate period structure observed in the longitudinal $\rho_{11}$, $\rho_{22}$ and transverse resistivity $\rho_{12}$ can be attributed to the interplay between symmetry and anisotropy of the Fermi surface geometry as the magnetic field rotates within the plane. It is also worth noting that the strength of the magnetic field plays a significant role in this interplay, thereby influencing the periodicity. For instance,
in Fig.\ref{fig4} (g) and (h), when we examine the resistivity components $\rho_{\parallel}$ and $\rho_{\perp}$, a six-fold symmetry becomes apparent. This symmetry arises because the projection of the Fermi surface onto the plane perpendicular to the current direction has $C_3$ symmetry. Therefore, both $\rho_{\parallel}$ and $\rho_{\perp}$ exhibit this invariance, reflecting the six-fold symmetric nature of the Fermi surface geometry.

Before discussing the details of planar Hall resistivity, it is crucial to highlight the difference in AMR between ferromagnetic materials and bismuth when applying the rotation transformation, a detail that can be easily overlooked. To illustrate this point, we assume that the rotation transformation of Eq.~\ref{rhoph} is correct to calculate the PHE of bismuth. With defining the longitudinal and transverse directions as the parallel and perpendicular axes respectively, we insert $\rho_{\parallel}$ and $\rho_{\perp}$ into $\widetilde{\rho}_{12}$ defined as $ \widetilde{\rho}_{12} \equiv  - ( \rho_{\perp} -\rho_{\parallel}) sin \theta cos \theta$~\footnote{Here we use $\widetilde{\rho}_{12}$(plotted in Fig.4(i)) to represent the planar Hall resistivity derived from the transformation of Eq.(7) for bismuth, in order to distinguish it from the one ($\rho_{12}$) calculated using Eq.(9) as shown in Fig.4(f).}. After plotting $\widetilde{\rho}_{12}$ in Fig.\ref{fig4} (i), however, it does not yield the same results as our direct calculations shown in Fig.\ref{fig4}(f), which accurately replicate the experimental measurements in Fig.\ref{fig4} (c). The discrepancy arises because the traditional rotation transformation in Eq.\ref{rhoph}, commonly used for ferromagnetic materials, considers only the diagonal elements of the resistivity tensor, while completely overlooking the off-diagonal elements.

Therefore, it becomes essential to incorporate a rotation transformation that involves all elements of the resistivity tensor, especially for materials with a highly anisotropic Fermi surface, as follows(see detailed derivation in ~\cite{supp}), 
\scriptsize 
\begin{widetext}
\begin{align}
\left(\begin{matrix}
  \frac{1}{2}[\rho_{\parallel} + \rho_{\perp}+ (\rho_{\parallel} - \rho_{\perp})\cos2\theta - (\rho_{x^{\prime}y^{\prime}} + \rho_{y^{\prime}x^{\prime}} )\sin2\theta] 
  & 
  \frac{1}{2}[\rho_{x^{\prime}y^{\prime}} - \rho_{y^{\prime}x^{\prime}} + (\rho_{x^{\prime}y^{\prime}} + \rho_{y^{\prime}x^{\prime}})\cos2\theta + (\rho_{\parallel} - \rho_{\perp})\sin2\theta]  
  & 
  \rho_{x^{\prime}z^{\prime}} \cos\theta -\rho_{y^{\prime}z^{\prime}} \sin\theta   \\    
  \frac{1}{2}[-\rho_{x^{\prime}y^{\prime}} + \rho_{y^{\prime}x^{\prime}} + (\rho_{x^{\prime}y^{\prime}} + \rho_{y^{\prime}x^{\prime}})\cos2\theta + (\rho_{\parallel} - \rho_{\perp})\sin2\theta] 
  & 
   \frac{1}{2}[\rho_{\parallel} + \rho_{\perp}+ (-\rho_{\parallel} + \rho_{\perp})\cos2\theta - (\rho_{x^{\prime}y^{\prime}} + \rho_{y^{\prime}x^{\prime}} )\sin2\theta]
   & 
   \rho_{y^{\prime}z^{\prime}} \cos\theta +\rho_{x^{\prime}z^{\prime}} \sin\theta   \\
  \rho_{z^{\prime}x^{\prime}} \cos\theta -\rho_{z^{\prime}y^{\prime}} \sin\theta
  & \rho_{z^{\prime}y^{\prime}} \cos\theta +\rho_{z^{\prime}x^{\prime}} \sin\theta 
  & \rho_{z^{\prime}z^{\prime}} \label{fullrotation}
\end{matrix}\right)
\end{align}
\end{widetext}
\normalsize
The difference in the rotation transformation between Eq.\ref{rhoph} and Eq.\ref{fullrotation} stems from the distinct AMR origins in ferromagnetic materials and bismuth. In ferromagnetic materials, an easy magnetization axis aligns with the external magnetic field, leading to significantly larger diagonal elements compared to off-diagonal elements in the resistivity tensor. Therefore, it is justified to use Eq.\ref{rhoxx} and Eq.\ref{rhoph} for the rotation transformation, disregarding the off-diagonal resistivity elements in these materials. In contrast, due to bismuth’s highly anisotropic Fermi surface, both diagonal ($\rho_{\parallel}$, $\rho_{\perp}$) and off-diagonal resistivity elements (in Eq.\ref{fullrotation}) are comparably significant. Consequently, a full rotation transformation is necessary to accurately replicate results, as illustrated in Fig.\ref{fig4}(f) and (i). 

With the full rotation of the resistivity tensor as shown in Eq.\ref{fullrotation}, we can gain a deeper understanding of the fine period structure, we examine the individual terms contributing to the resistivity element $\rho_{11}$, which is expressed as $\rho_{11} = \frac{1}{2}[(\rho_{\parallel} + \rho_{\perp})+ (\rho_{\parallel} - \rho_{\perp})\cos2\theta - (\rho_{x'y'} + \rho_{y'x'})\sin2\theta]$. By analyzing these terms one by one, one observes the following features. At low magnetic field ($B\tau = 0.8 \rm T\cdot ps$), the third term, involving $(\rho_{x'y'} + \rho_{y'x'})\sin2\theta$, displays a period of $\frac{\pi}{2}$, as shown in Fig.S1~\cite{supp}. This term predominantly influences the behavior of $\rho_{11}$ in the low magnetic field regime. Conversely, at high magnetic field ($B\tau  = 3.2 \rm T\cdot ps$), the first and second terms, containing $\rho_{\parallel}$ and $\rho_{\perp}$,  which are nearly identical as demonstrated in Fig.S1~\cite{supp}, become the principal contributors to $\rho_{11}$. The distinction in resistivity behavior under different magnetic field limits is quite logical, given that $\rho_{\parallel}$ and $\rho_{\perp}$ vary quadratically with the magnetic field (as $B^2$), while $\rho_{xy}$ and $\rho_{yx}$ change linearly with $B$. As a result, at lower magnetic field, the third term comprising $\rho_{xy}$ and $\rho_{yx}$ predominantly influences the period structure. Conversely, at higher magnetic field, the first and second terms, associated with $\rho_{\parallel}$ and $\rho_{\perp}$ respectively, become the key determinants.

In conclusion, the PHE is not a transport phenomenon exclusive to ferromagnetic and nontrivial topological materials; it also occurs in ordinary materials, such as bismuth. While the PHE in both ferromagnetic materials and bismuth arises from AMR, their underlying causes differ. In ferromagnetic materials, the PHE is mainly attributed to spin-orbit coupling.  Whereas in bismuth, it results from the anisotropic Fermi surface geometry. This distinction necessitates a rigorous approach when applying rotation transformation to bismuth, requiring the inclusion of all resistivity tensor elements, not just the diagonal ones, as is typically done for ferromagnetic materials. In all, understanding the varying origins and mechanisms of PHE across different materials not only sheds light on the multifaceted nature of this phenomenon but also enhances our overall grasp of the underlying physics in diverse material systems. 

\section{Hall effects with Berry curvature}
When Berry curvature(BC) is present, the equation of motion incorporates an additional anomalous velocity term ~\cite{Chang1996}. In systems where time-reversal symmetry is broken, this anomalous velocity gives rise to Hall conductivity. Conversely, in systems with time-reversal symmetry, symmetry constraints negate the anomalous Hall effect. A magnetic field can induce BC and, consequently, the anomalous Hall effect. For non-magnetic materials, the induced anomalous Hall effect is typically minimal and negligible compared to the conductivity from a larger Fermi surface. However, in materials with low carrier concentrations, where conductivity contributions are small, the anomalous Hall effect driven by magnetically induced Berry curvature becomes significant. While this topic has been extensively studied, it is not the primary focus of our research. We shall explore the Hall effect in magnetic materials to examine the impact of the anomalous Hall effect caused by Berry curvature on the overall Hall effect. The intrinsic BC of magnetic materials or that induced by a magnetic field is significant and substantially influences the Hall effect. 

Although the calculation of magnetic transport properties in magnetic materials has been addressed in an another work~\cite{liu2024}, for completeness, we will briefly revisit the method here. Taking into account both the Lorentz force and the Hall effect due to Berry curvature, the total conductance is given by,

\begin{equation}
    {\bm \sigma}=\sum_{n}{\bm \sigma}_n^O(\bm B, T)+\bm \sigma^A(\bm B,T),\label{sigma_total}
\end{equation}
where $n$ denotes the band index, ${\bm \sigma}_n^O(\bm B, T)$ refers to the ordinary conductivity of the $n$-th band, which results from the combined effects of the Lorentz force and scattering processes, and $\bm \sigma^A(\bm B, T)$ represents the AHC that originates from system's magnetic properties.
\begin{gather}
    \begin{split}(\sigma_{\alpha\beta}^A)_s&=-\varepsilon_{\alpha\beta\gamma}\frac{e^2}{\hbar}\sum_n\int_{\rm BZ}\frac{\diff^3 \bm k}{(2\pi)^3}f(\epsilon_n(\bm k))\Omega_n^{\gamma}(\bm k).\label{sigma_ahc_s}
    \end{split}
\end{gather}
$(\sigma_{\alpha\beta}^A)_s$ represents the maximum intrinsic AHC when the magnetization is fully saturated and oriented perpendicular to the $\alpha$-$\beta$ plane. This magnetization can arise either spontaneously in ferromagnetic materials or be induced externally in paramagnetic and antiferromagnetic materials. Although the AHC may exhibit a nonlinear dependency on magnetization, we assume that the AHC is proportional to the magnetization in magnets, as suggested by \cite{zeng2006linear}. When the magnetization is aligned parallel to the magnetic field in the $\hat{z}$ direction, the AHC can be described by:
\begin{gather}
    \sigma_{xy}^A(B,T)=\frac{M(B,T)}{M_s}       (\sigma_{xy}^A)_{s},\label{sigmaA}
\end{gather}
Here, $M(B,T)$ represents the magnetization that depends on both the magnetic field and temperature, while $M_s$ denotes the saturated magnetization. In this study, for the sake of simplicity, we focus on the single-band scenario, where the conductivity can be described as follows ~\cite{zhao2023magnetotransport}:
\begin{gather}
    \bm \sigma=\begin{bmatrix}
        \sigma_{xx}^O & \sigma_{xy}^O+\sigma_{xy}^A\\
        \sigma_{yx}^O-\sigma_{xy}^A & \sigma_{yy}^O
    \end{bmatrix},
\end{gather}
where $\sigma_{xx}^O$, $\sigma_{yy}^O$ and $\sigma_{yx}^O$ are the ordinary conductivities. 
The Hall resistivity is obtained by taking the inverse of the conductivity tensor $\bm \sigma$, and can be expressed as follows:
\begin{gather}
    \rho_{yx}=\frac{\sigma_{xy}^O+\sigma_{xy}^A}{\sigma_{xx}\sigma_{yy}+(\sigma_{xy}^O+\sigma_{xy}^A)^2}.
\end{gather}

According to Eq.(\ref{sigmaA}), magnetic field and temperature alter the magnetization curve, which affects the anomalous Hall conductance (AHC) and hence influences the Hall effect. In this work, we focus on the simple linear magnetization curve, i.e., $M \propto B$, typical of paramagnetic or antiferromagnetic materials  as sketched in Fig.\ref{fig5}a. For an extensive discussion on other scenarios, please see Ref.\cite{liu2024}. Additionally, it is important to note that the signs of the conventional Hall coefficient are determined by the type of carriers, while the sign of the anomalous Hall effect is dictated by the BC.

The results are displayed in Fig.\ref{fig5}b. At high temperatures, where carrier mobility is tinny, both the longitudinal conductance, $\sigma_{xx}$, and the normal Hall conductance, $\sigma_{xy}^O$, are significantly reduced. Under these conditions, the Hall resistance is predominantly influenced by the AHE, and the slope of the Hall curve is dictated by the sign of the AHC. Conversely, at low temperatures, carrier mobility increases substantially, and the conductance resulting from the Lorentz force significantly outweighs that of the AHC, thus making the Hall coefficient dependent on the carrier type. When the signs of the AHC and the normal Hall coefficient are opposite, an interesting phenomenon occurs: the sign of the Hall coefficient changes with temperature.

\begin{figure}
    \centering
\includegraphics[width=0.45\textwidth]{ 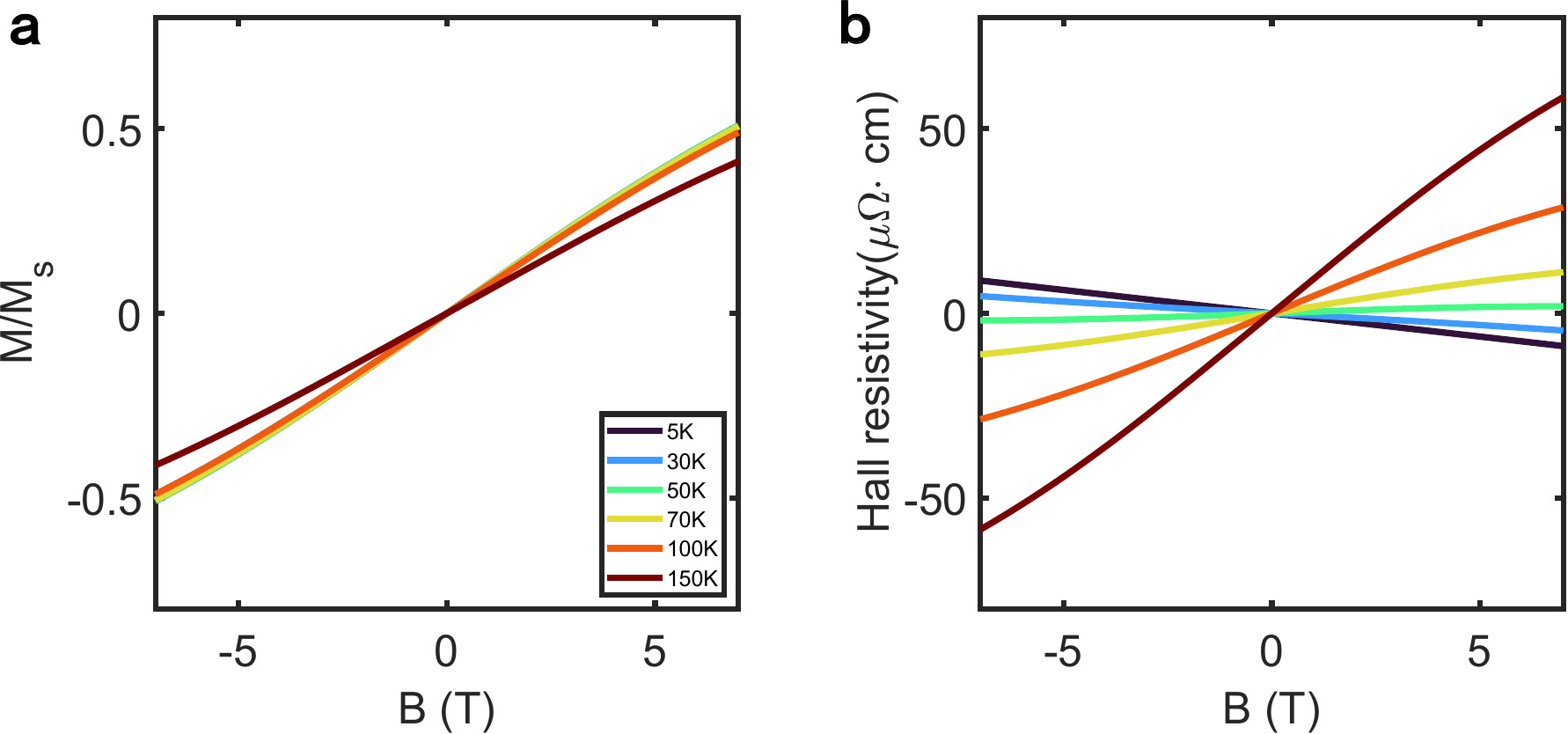}
    \caption{Field-dependent  (a) magnetization and (b) Hall resistivity at different temperatures.}
\label{fig5}
\end{figure}

\section{Conclusion}
In conclusion, we advanced a first-principles Boltzmann transport methodology to systematically investigate the complex behavior of the Hall effect, focusing on its dependency on magnetic fields, temperature, and angles. We elucidated that non-linear Hall curves in non-magnetic materials, similar to the AHE in magnetic materials, originate from the transport behavior involving multiple types of charge carriers, which is determined by the intrinsic geometry of the material's Fermi surface. We proposed a scaling behavior for Hall curves, analogous to Kohler's rule, which accounts for temperature and magnetic field variations, explaining the temperature-dependent behaviors of Hall curves. Ultimately, our method successfully explained the planar Hall effect observed in bismuth. This research highlights the critical role of a material's intrinsic Fermi surface geometry, including size and other properties, in shaping the Hall response of non-magnetic materials. Our proposed approach allows for the study of Hall effects from first-principles, and conversely, experimental measurements of the Hall effect can reveal intrinsic electronic structural properties of materials. Finally, we also discussed the influence of Berry curvature in magnetic materials on the Hall effect. We found that the presence of Berry curvature can result in the interesting phenomenon of Hall sign reversal with temperature in certain systems.
Our work marks a significant shift, treating Hall resistivity scaling as an intrinsic material trait and offering a broader framework for identifying and characterizing materials. This advancement implies a potential paradigm change in material science, steering away from extrinsic-dependent explanations to intrinsic material properties.

\section{Acknowledgments}
We acknowledge the invaluable discussions with X. Dai. This work was supported by the National Key R\&D Program of China (Grant No. 2023YFA1607400, 2022YFA1403800), the National Natural Science Foundation of China (Grant No.12274436, 11925408, 11921004), the Science Center of the National Natural Science Foundation of China (Grant No. 12188101), and  H.W. acknowledge support from the Informatization Plan of the Chinese Academy of Sciences (CASWX2021SF-0102) and the New Cornerstone Science Foundation through the XPLORER PRIZE.

\bibliography{refs}

\begin{thebibliography}{73}%
\makeatletter
\providecommand \@ifxundefined [1]{%
 \@ifx{#1\undefined}
}%
\providecommand \@ifnum [1]{%
 \ifnum #1\expandafter \@firstoftwo
 \else \expandafter \@secondoftwo
 \fi
}%
\providecommand \@ifx [1]{%
 \ifx #1\expandafter \@firstoftwo
 \else \expandafter \@secondoftwo
 \fi
}%
\providecommand \natexlab [1]{#1}%
\providecommand \enquote  [1]{``#1''}%
\providecommand \bibnamefont  [1]{#1}%
\providecommand \bibfnamefont [1]{#1}%
\providecommand \citenamefont [1]{#1}%
\providecommand \href@noop [0]{\@secondoftwo}%
\providecommand \href [0]{\begingroup \@sanitize@url \@href}%
\providecommand \@href[1]{\@@startlink{#1}\@@href}%
\providecommand \@@href[1]{\endgroup#1\@@endlink}%
\providecommand \@sanitize@url [0]{\catcode `\\12\catcode `\$12\catcode
  `\&12\catcode `\#12\catcode `\^12\catcode `\_12\catcode `\%12\relax}%
\providecommand \@@startlink[1]{}%
\providecommand \@@endlink[0]{}%
\providecommand \url  [0]{\begingroup\@sanitize@url \@url }%
\providecommand \@url [1]{\endgroup\@href {#1}{\urlprefix }}%
\providecommand \urlprefix  [0]{URL }%
\providecommand \Eprint [0]{\href }%
\providecommand \doibase [0]{http://dx.doi.org/}%
\providecommand \selectlanguage [0]{\@gobble}%
\providecommand \bibinfo  [0]{\@secondoftwo}%
\providecommand \bibfield  [0]{\@secondoftwo}%
\providecommand \translation [1]{[#1]}%
\providecommand \BibitemOpen [0]{}%
\providecommand \bibitemStop [0]{}%
\providecommand \bibitemNoStop [0]{.\EOS\space}%
\providecommand \EOS [0]{\spacefactor3000\relax}%
\providecommand \BibitemShut  [1]{\csname bibitem#1\endcsname}%
\let\auto@bib@innerbib\@empty
\bibitem [{\citenamefont {Hall}(1879)}]{Hall1879}%
  \BibitemOpen
  \bibfield  {author} {\bibinfo {author} {\bibfnamefont {E.~H.}\ \bibnamefont
  {Hall}},\ }\href {http://www.jstor.org/stable/2369245} {\bibfield  {journal}
  {\bibinfo  {journal} {American Journal of Mathematics}\ }\textbf {\bibinfo
  {volume} {2}},\ \bibinfo {pages} {287} (\bibinfo {year} {1879})}\BibitemShut
  {NoStop}%
\bibitem [{\citenamefont {Hall}(1881)}]{Hall1881}%
  \BibitemOpen
  \bibfield  {author} {\bibinfo {author} {\bibfnamefont {E.~H.}\ \bibnamefont
  {Hall}},\ }\href@noop {} {\bibfield  {journal} {\bibinfo  {journal} {Philos.
  Mag.}\ }\textbf {\bibinfo {volume} {12}},\ \bibinfo {pages} {157} (\bibinfo
  {year} {1881})}\BibitemShut {NoStop}%
\bibitem [{\citenamefont {Nagaosa}\ \emph {et~al.}(2010)\citenamefont
  {Nagaosa}, \citenamefont {Sinova}, \citenamefont {Onoda}, \citenamefont
  {MacDonald},\ and\ \citenamefont {Ong}}]{AHERMP}%
  \BibitemOpen
  \bibfield  {author} {\bibinfo {author} {\bibfnamefont {N.}~\bibnamefont
  {Nagaosa}}, \bibinfo {author} {\bibfnamefont {J.}~\bibnamefont {Sinova}},
  \bibinfo {author} {\bibfnamefont {S.}~\bibnamefont {Onoda}}, \bibinfo
  {author} {\bibfnamefont {A.~H.}\ \bibnamefont {MacDonald}}, \ and\ \bibinfo
  {author} {\bibfnamefont {N.~P.}\ \bibnamefont {Ong}},\ }\href {\doibase
  10.1103/RevModPhys.82.1539} {\bibfield  {journal} {\bibinfo  {journal} {Rev.
  Mod. Phys.}\ }\textbf {\bibinfo {volume} {82}},\ \bibinfo {pages} {1539}
  (\bibinfo {year} {2010})}\BibitemShut {NoStop}%
\bibitem [{\citenamefont {Klitzing}\ \emph {et~al.}(1980)\citenamefont
  {Klitzing}, \citenamefont {Dorda},\ and\ \citenamefont
  {Pepper}}]{KlitzingQHE}%
  \BibitemOpen
  \bibfield  {author} {\bibinfo {author} {\bibfnamefont {K.~v.}\ \bibnamefont
  {Klitzing}}, \bibinfo {author} {\bibfnamefont {G.}~\bibnamefont {Dorda}}, \
  and\ \bibinfo {author} {\bibfnamefont {M.}~\bibnamefont {Pepper}},\ }\href
  {\doibase 10.1103/PhysRevLett.45.494} {\bibfield  {journal} {\bibinfo
  {journal} {Phys. Rev. Lett.}\ }\textbf {\bibinfo {volume} {45}},\ \bibinfo
  {pages} {494} (\bibinfo {year} {1980})}\BibitemShut {NoStop}%
\bibitem [{\citenamefont {Dyakonov}\ and\ \citenamefont
  {Perel}(1971{\natexlab{a}})}]{SHEJEPT}%
  \BibitemOpen
  \bibfield  {author} {\bibinfo {author} {\bibfnamefont {M.~I.}\ \bibnamefont
  {Dyakonov}}\ and\ \bibinfo {author} {\bibfnamefont {V.~I.}\ \bibnamefont
  {Perel}},\ }\href@noop {} {\bibfield  {journal} {\bibinfo  {journal} {Sov.
  Phys. JETP Lett.}\ }\textbf {\bibinfo {volume} {13}},\ \bibinfo {pages} {467}
  (\bibinfo {year} {1971}{\natexlab{a}})}\BibitemShut {NoStop}%
\bibitem [{\citenamefont {Dyakonov}\ and\ \citenamefont
  {Perel}(1971{\natexlab{b}})}]{SHEPLA}%
  \BibitemOpen
  \bibfield  {author} {\bibinfo {author} {\bibfnamefont {M.}~\bibnamefont
  {Dyakonov}}\ and\ \bibinfo {author} {\bibfnamefont {V.}~\bibnamefont
  {Perel}},\ }\href {\doibase https://doi.org/10.1016/0375-9601(71)90196-4}
  {\bibfield  {journal} {\bibinfo  {journal} {Physics Letters A}\ }\textbf
  {\bibinfo {volume} {35}},\ \bibinfo {pages} {459} (\bibinfo {year}
  {1971}{\natexlab{b}})}\BibitemShut {NoStop}%
\bibitem [{\citenamefont {Hong}\ and\ \citenamefont {Giordano}(1995)}]{PHEprb}%
  \BibitemOpen
  \bibfield  {author} {\bibinfo {author} {\bibfnamefont {K.}~\bibnamefont
  {Hong}}\ and\ \bibinfo {author} {\bibfnamefont {N.}~\bibnamefont
  {Giordano}},\ }\href {\doibase 10.1103/PhysRevB.51.9855} {\bibfield
  {journal} {\bibinfo  {journal} {Phys. Rev. B}\ }\textbf {\bibinfo {volume}
  {51}},\ \bibinfo {pages} {9855} (\bibinfo {year} {1995})}\BibitemShut
  {NoStop}%
\bibitem [{\citenamefont {Ohno}(1998)}]{PHEscience}%
  \BibitemOpen
  \bibfield  {author} {\bibinfo {author} {\bibfnamefont {H.}~\bibnamefont
  {Ohno}},\ }\href {\doibase 10.1126/science.281.5379.951} {\bibfield
  {journal} {\bibinfo  {journal} {Science}\ }\textbf {\bibinfo {volume}
  {281}},\ \bibinfo {pages} {951} (\bibinfo {year} {1998})}\BibitemShut
  {NoStop}%
\bibitem [{\citenamefont {Tang}\ \emph {et~al.}(2003)\citenamefont {Tang},
  \citenamefont {Kawakami}, \citenamefont {Awschalom},\ and\ \citenamefont
  {Roukes}}]{PHEGaMnAs}%
  \BibitemOpen
  \bibfield  {author} {\bibinfo {author} {\bibfnamefont {H.~X.}\ \bibnamefont
  {Tang}}, \bibinfo {author} {\bibfnamefont {R.~K.}\ \bibnamefont {Kawakami}},
  \bibinfo {author} {\bibfnamefont {D.~D.}\ \bibnamefont {Awschalom}}, \ and\
  \bibinfo {author} {\bibfnamefont {M.~L.}\ \bibnamefont {Roukes}},\ }\href
  {\doibase 10.1103/PhysRevLett.90.107201} {\bibfield  {journal} {\bibinfo
  {journal} {Phys. Rev. Lett.}\ }\textbf {\bibinfo {volume} {90}},\ \bibinfo
  {pages} {107201} (\bibinfo {year} {2003})}\BibitemShut {NoStop}%
\bibitem [{\citenamefont {Li}\ \emph {et~al.}(2016{\natexlab{a}})\citenamefont
  {Li}, \citenamefont {Kharzeev}, \citenamefont {Zhang}, \citenamefont {Huang},
  \citenamefont {Pletikosi{\'{c}}}, \citenamefont {Fedorov}, \citenamefont
  {Zhong}, \citenamefont {Schneeloch}, \citenamefont {Gu},\ and\ \citenamefont
  {Valla}}]{NMRnaturephy}%
  \BibitemOpen
  \bibfield  {author} {\bibinfo {author} {\bibfnamefont {Q.}~\bibnamefont
  {Li}}, \bibinfo {author} {\bibfnamefont {D.~E.}\ \bibnamefont {Kharzeev}},
  \bibinfo {author} {\bibfnamefont {C.}~\bibnamefont {Zhang}}, \bibinfo
  {author} {\bibfnamefont {Y.}~\bibnamefont {Huang}}, \bibinfo {author}
  {\bibfnamefont {I.}~\bibnamefont {Pletikosi{\'{c}}}}, \bibinfo {author}
  {\bibfnamefont {A.}~\bibnamefont {Fedorov}}, \bibinfo {author} {\bibfnamefont
  {R.}~\bibnamefont {Zhong}}, \bibinfo {author} {\bibfnamefont
  {J.}~\bibnamefont {Schneeloch}}, \bibinfo {author} {\bibfnamefont
  {G.}~\bibnamefont {Gu}}, \ and\ \bibinfo {author} {\bibfnamefont
  {T.}~\bibnamefont {Valla}},\ }\href {\doibase 10.1038/nphys3648} {\bibfield
  {journal} {\bibinfo  {journal} {Nature Physics}\ }\textbf {\bibinfo {volume}
  {12}},\ \bibinfo {pages} {550} (\bibinfo {year}
  {2016}{\natexlab{a}})}\BibitemShut {NoStop}%
\bibitem [{\citenamefont {Adler}(1969)}]{CApr}%
  \BibitemOpen
  \bibfield  {author} {\bibinfo {author} {\bibfnamefont {S.~L.}\ \bibnamefont
  {Adler}},\ }\href {\doibase 10.1103/PhysRev.177.2426} {\bibfield  {journal}
  {\bibinfo  {journal} {Phys. Rev.}\ }\textbf {\bibinfo {volume} {177}},\
  \bibinfo {pages} {2426} (\bibinfo {year} {1969})}\BibitemShut {NoStop}%
\bibitem [{\citenamefont {Bell}\ and\ \citenamefont
  {Jackiw}(1969)}]{CANuovoCim}%
  \BibitemOpen
  \bibfield  {author} {\bibinfo {author} {\bibfnamefont {J.~S.}\ \bibnamefont
  {Bell}}\ and\ \bibinfo {author} {\bibfnamefont {R.}~\bibnamefont {Jackiw}},\
  }\href {\doibase 10.1007/BF02823296} {\bibfield  {journal} {\bibinfo
  {journal} {Il Nuovo Cimento A (1965-1970)}\ }\textbf {\bibinfo {volume}
  {60}},\ \bibinfo {pages} {47} (\bibinfo {year} {1969})}\BibitemShut {NoStop}%
\bibitem [{\citenamefont {Maryenko}\ \emph {et~al.}(2017)\citenamefont
  {Maryenko}, \citenamefont {Mishchenko}, \citenamefont {Bahramy},
  \citenamefont {Ernst}, \citenamefont {Falson}, \citenamefont {Kozuka},
  \citenamefont {Tsukazaki}, \citenamefont {Nagaosa},\ and\ \citenamefont
  {Kawasaki}}]{Maryenko2017}%
  \BibitemOpen
  \bibfield  {author} {\bibinfo {author} {\bibfnamefont {D.}~\bibnamefont
  {Maryenko}}, \bibinfo {author} {\bibfnamefont {A.~S.}\ \bibnamefont
  {Mishchenko}}, \bibinfo {author} {\bibfnamefont {M.~S.}\ \bibnamefont
  {Bahramy}}, \bibinfo {author} {\bibfnamefont {A.}~\bibnamefont {Ernst}},
  \bibinfo {author} {\bibfnamefont {J.}~\bibnamefont {Falson}}, \bibinfo
  {author} {\bibfnamefont {Y.}~\bibnamefont {Kozuka}}, \bibinfo {author}
  {\bibfnamefont {A.}~\bibnamefont {Tsukazaki}}, \bibinfo {author}
  {\bibfnamefont {N.}~\bibnamefont {Nagaosa}}, \ and\ \bibinfo {author}
  {\bibfnamefont {M.}~\bibnamefont {Kawasaki}},\ }\href {\doibase
  10.1038/ncomms14777} {\bibfield  {journal} {\bibinfo  {journal} {Nature
  Communications}\ }\textbf {\bibinfo {volume} {8}} (\bibinfo {year} {2017}),\
  10.1038/ncomms14777}\BibitemShut {NoStop}%
\bibitem [{\citenamefont {Liang}\ \emph {et~al.}(2018)\citenamefont {Liang},
  \citenamefont {Lin}, \citenamefont {Gibson}, \citenamefont {Kushwaha},
  \citenamefont {Liu}, \citenamefont {Wang}, \citenamefont {Xiong},
  \citenamefont {Sobota}, \citenamefont {Hashimoto}, \citenamefont {Kirchmann},
  \citenamefont {Shen}, \citenamefont {Cava},\ and\ \citenamefont
  {Ong}}]{AHEZrTe5Liang}%
  \BibitemOpen
  \bibfield  {author} {\bibinfo {author} {\bibfnamefont {T.}~\bibnamefont
  {Liang}}, \bibinfo {author} {\bibfnamefont {J.}~\bibnamefont {Lin}}, \bibinfo
  {author} {\bibfnamefont {Q.}~\bibnamefont {Gibson}}, \bibinfo {author}
  {\bibfnamefont {S.}~\bibnamefont {Kushwaha}}, \bibinfo {author}
  {\bibfnamefont {M.}~\bibnamefont {Liu}}, \bibinfo {author} {\bibfnamefont
  {W.}~\bibnamefont {Wang}}, \bibinfo {author} {\bibfnamefont {H.}~\bibnamefont
  {Xiong}}, \bibinfo {author} {\bibfnamefont {J.~A.}\ \bibnamefont {Sobota}},
  \bibinfo {author} {\bibfnamefont {M.}~\bibnamefont {Hashimoto}}, \bibinfo
  {author} {\bibfnamefont {P.~S.}\ \bibnamefont {Kirchmann}}, \bibinfo {author}
  {\bibfnamefont {Z.-X.}\ \bibnamefont {Shen}}, \bibinfo {author}
  {\bibfnamefont {R.~J.}\ \bibnamefont {Cava}}, \ and\ \bibinfo {author}
  {\bibfnamefont {N.~P.}\ \bibnamefont {Ong}},\ }\href {\doibase
  10.1038/s41567-018-0078-z} {\bibfield  {journal} {\bibinfo  {journal} {Nature
  Physics}\ }\textbf {\bibinfo {volume} {14}},\ \bibinfo {pages} {451}
  (\bibinfo {year} {2018})}\BibitemShut {NoStop}%
\bibitem [{\citenamefont {Chang}\ and\ \citenamefont {Niu}(1996)}]{Chang1996}%
  \BibitemOpen
  \bibfield  {author} {\bibinfo {author} {\bibfnamefont {M.-C.}\ \bibnamefont
  {Chang}}\ and\ \bibinfo {author} {\bibfnamefont {Q.}~\bibnamefont {Niu}},\
  }\href {\doibase 10.1103/PhysRevB.53.7010} {\bibfield  {journal} {\bibinfo
  {journal} {Phys. Rev. B}\ }\textbf {\bibinfo {volume} {53}},\ \bibinfo
  {pages} {7010} (\bibinfo {year} {1996})}\BibitemShut {NoStop}%
\bibitem [{\citenamefont {Sundaram}\ and\ \citenamefont
  {Niu}(1999)}]{Sundaram1999}%
  \BibitemOpen
  \bibfield  {author} {\bibinfo {author} {\bibfnamefont {G.}~\bibnamefont
  {Sundaram}}\ and\ \bibinfo {author} {\bibfnamefont {Q.}~\bibnamefont {Niu}},\
  }\href {\doibase 10.1103/PhysRevB.59.14915} {\bibfield  {journal} {\bibinfo
  {journal} {Phys. Rev. B}\ }\textbf {\bibinfo {volume} {59}},\ \bibinfo
  {pages} {14915} (\bibinfo {year} {1999})}\BibitemShut {NoStop}%
\bibitem [{\citenamefont {Arno}\ \emph {et~al.}(2003)\citenamefont {Arno},
  \citenamefont {Ali}, \citenamefont {Hiroyasu}, \citenamefont {Qian},\ and\
  \citenamefont {Josef}}]{Bohm2003}%
  \BibitemOpen
  \bibfield  {author} {\bibinfo {author} {\bibfnamefont {B.}~\bibnamefont
  {Arno}}, \bibinfo {author} {\bibfnamefont {M.}~\bibnamefont {Ali}}, \bibinfo
  {author} {\bibfnamefont {K.}~\bibnamefont {Hiroyasu}}, \bibinfo {author}
  {\bibfnamefont {N.}~\bibnamefont {Qian}}, \ and\ \bibinfo {author}
  {\bibfnamefont {Z.}~\bibnamefont {Josef}},\ }\href@noop {} {\emph {\bibinfo
  {title} {The Geometric Phase in Quantum Systems}}}\ (\bibinfo  {publisher}
  {Springer, Berlin},\ \bibinfo {year} {2003})\BibitemShut {NoStop}%
\bibitem [{\citenamefont {Xiao}\ \emph {et~al.}(2010)\citenamefont {Xiao},
  \citenamefont {Chang},\ and\ \citenamefont {Niu}}]{XiaoDiRMP}%
  \BibitemOpen
  \bibfield  {author} {\bibinfo {author} {\bibfnamefont {D.}~\bibnamefont
  {Xiao}}, \bibinfo {author} {\bibfnamefont {M.-C.}\ \bibnamefont {Chang}}, \
  and\ \bibinfo {author} {\bibfnamefont {Q.}~\bibnamefont {Niu}},\ }\href
  {\doibase 10.1103/RevModPhys.82.1959} {\bibfield  {journal} {\bibinfo
  {journal} {Rev. Mod. Phys.}\ }\textbf {\bibinfo {volume} {82}},\ \bibinfo
  {pages} {1959} (\bibinfo {year} {2010})}\BibitemShut {NoStop}%
\bibitem [{\citenamefont {Majumdar}\ and\ \citenamefont
  {Berger}(1973)}]{AHEFeprb73}%
  \BibitemOpen
  \bibfield  {author} {\bibinfo {author} {\bibfnamefont {A.~K.}\ \bibnamefont
  {Majumdar}}\ and\ \bibinfo {author} {\bibfnamefont {L.}~\bibnamefont
  {Berger}},\ }\href {\doibase 10.1103/PhysRevB.7.4203} {\bibfield  {journal}
  {\bibinfo  {journal} {Phys. Rev. B}\ }\textbf {\bibinfo {volume} {7}},\
  \bibinfo {pages} {4203} (\bibinfo {year} {1973})}\BibitemShut {NoStop}%
\bibitem [{\citenamefont {Schad}\ \emph {et~al.}(1998)\citenamefont {Schad},
  \citenamefont {Beliën}, \citenamefont {Verbanck}, \citenamefont
  {Moshchalkov},\ and\ \citenamefont {Bruynseraede}}]{Schadjpcm1998}%
  \BibitemOpen
  \bibfield  {author} {\bibinfo {author} {\bibfnamefont {R.}~\bibnamefont
  {Schad}}, \bibinfo {author} {\bibfnamefont {P.}~\bibnamefont {Beliën}},
  \bibinfo {author} {\bibfnamefont {G.}~\bibnamefont {Verbanck}}, \bibinfo
  {author} {\bibfnamefont {V.~V.}\ \bibnamefont {Moshchalkov}}, \ and\ \bibinfo
  {author} {\bibfnamefont {Y.}~\bibnamefont {Bruynseraede}},\ }\href {\doibase
  10.1088/0953-8984/10/30/005} {\bibfield  {journal} {\bibinfo  {journal}
  {Journal of Physics: Condensed Matter}\ }\textbf {\bibinfo {volume} {10}},\
  \bibinfo {pages} {6643} (\bibinfo {year} {1998})}\BibitemShut {NoStop}%
\bibitem [{\citenamefont {Shiomi}\ \emph {et~al.}(2009)\citenamefont {Shiomi},
  \citenamefont {Onose},\ and\ \citenamefont {Tokura}}]{AHEFeprb09}%
  \BibitemOpen
  \bibfield  {author} {\bibinfo {author} {\bibfnamefont {Y.}~\bibnamefont
  {Shiomi}}, \bibinfo {author} {\bibfnamefont {Y.}~\bibnamefont {Onose}}, \
  and\ \bibinfo {author} {\bibfnamefont {Y.}~\bibnamefont {Tokura}},\ }\href
  {\doibase 10.1103/PhysRevB.79.100404} {\bibfield  {journal} {\bibinfo
  {journal} {Phys. Rev. B}\ }\textbf {\bibinfo {volume} {79}},\ \bibinfo
  {pages} {100404} (\bibinfo {year} {2009})}\BibitemShut {NoStop}%
\bibitem [{\citenamefont {Bowen}\ \emph {et~al.}(2005)\citenamefont {Bowen},
  \citenamefont {Friedland}, \citenamefont {Herfort}, \citenamefont
  {Sch\"onherr},\ and\ \citenamefont {Ploog}}]{PHEFe3Si2005}%
  \BibitemOpen
  \bibfield  {author} {\bibinfo {author} {\bibfnamefont {M.}~\bibnamefont
  {Bowen}}, \bibinfo {author} {\bibfnamefont {K.-J.}\ \bibnamefont
  {Friedland}}, \bibinfo {author} {\bibfnamefont {J.}~\bibnamefont {Herfort}},
  \bibinfo {author} {\bibfnamefont {H.-P.}\ \bibnamefont {Sch\"onherr}}, \ and\
  \bibinfo {author} {\bibfnamefont {K.~H.}\ \bibnamefont {Ploog}},\ }\href
  {\doibase 10.1103/PhysRevB.71.172401} {\bibfield  {journal} {\bibinfo
  {journal} {Phys. Rev. B}\ }\textbf {\bibinfo {volume} {71}},\ \bibinfo
  {pages} {172401} (\bibinfo {year} {2005})}\BibitemShut {NoStop}%
\bibitem [{\citenamefont {Fernandez-Pacheco}\ \emph {et~al.}(2008)\citenamefont
  {Fernandez-Pacheco}, \citenamefont {De~Teresa}, \citenamefont {Orna},
  \citenamefont {Morellon}, \citenamefont {Algarabel}, \citenamefont {Pardo},
  \citenamefont {Ibarra}, \citenamefont {Magen},\ and\ \citenamefont
  {Snoeck}}]{PHEFe3O42008}%
  \BibitemOpen
  \bibfield  {author} {\bibinfo {author} {\bibfnamefont {A.}~\bibnamefont
  {Fernandez-Pacheco}}, \bibinfo {author} {\bibfnamefont {J.~M.}\ \bibnamefont
  {De~Teresa}}, \bibinfo {author} {\bibfnamefont {J.}~\bibnamefont {Orna}},
  \bibinfo {author} {\bibfnamefont {L.}~\bibnamefont {Morellon}}, \bibinfo
  {author} {\bibfnamefont {P.~A.}\ \bibnamefont {Algarabel}}, \bibinfo {author}
  {\bibfnamefont {J.~A.}\ \bibnamefont {Pardo}}, \bibinfo {author}
  {\bibfnamefont {M.~R.}\ \bibnamefont {Ibarra}}, \bibinfo {author}
  {\bibfnamefont {C.}~\bibnamefont {Magen}}, \ and\ \bibinfo {author}
  {\bibfnamefont {E.}~\bibnamefont {Snoeck}},\ }\href {\doibase
  10.1103/PhysRevB.78.212402} {\bibfield  {journal} {\bibinfo  {journal} {Phys.
  Rev. B}\ }\textbf {\bibinfo {volume} {78}},\ \bibinfo {pages} {212402}
  (\bibinfo {year} {2008})}\BibitemShut {NoStop}%
\bibitem [{\citenamefont {Xiong}\ \emph {et~al.}(2015)\citenamefont {Xiong},
  \citenamefont {Kushwaha}, \citenamefont {Liang}, \citenamefont {Krizan},
  \citenamefont {Hirschberger}, \citenamefont {Wang}, \citenamefont {Cava},\
  and\ \citenamefont {Ong}}]{PHENa3Bi}%
  \BibitemOpen
  \bibfield  {author} {\bibinfo {author} {\bibfnamefont {J.}~\bibnamefont
  {Xiong}}, \bibinfo {author} {\bibfnamefont {S.~K.}\ \bibnamefont {Kushwaha}},
  \bibinfo {author} {\bibfnamefont {T.}~\bibnamefont {Liang}}, \bibinfo
  {author} {\bibfnamefont {J.~W.}\ \bibnamefont {Krizan}}, \bibinfo {author}
  {\bibfnamefont {M.}~\bibnamefont {Hirschberger}}, \bibinfo {author}
  {\bibfnamefont {W.}~\bibnamefont {Wang}}, \bibinfo {author} {\bibfnamefont
  {R.~J.}\ \bibnamefont {Cava}}, \ and\ \bibinfo {author} {\bibfnamefont
  {N.~P.}\ \bibnamefont {Ong}},\ }\href {\doibase 10.1126/science.aac6089}
  {\bibfield  {journal} {\bibinfo  {journal} {Science}\ }\textbf {\bibinfo
  {volume} {350}},\ \bibinfo {pages} {413} (\bibinfo {year}
  {2015})}\BibitemShut {NoStop}%
\bibitem [{\citenamefont {Li}\ \emph {et~al.}(2016{\natexlab{b}})\citenamefont
  {Li}, \citenamefont {Kharzeev}, \citenamefont {Zhang}, \citenamefont {Huang},
  \citenamefont {Pletikosi{\'{c}}}, \citenamefont {Fedorov}, \citenamefont
  {Zhong}, \citenamefont {Schneeloch}, \citenamefont {Gu},\ and\ \citenamefont
  {Valla}}]{PHEZrTe5}%
  \BibitemOpen
  \bibfield  {author} {\bibinfo {author} {\bibfnamefont {Q.}~\bibnamefont
  {Li}}, \bibinfo {author} {\bibfnamefont {D.~E.}\ \bibnamefont {Kharzeev}},
  \bibinfo {author} {\bibfnamefont {C.}~\bibnamefont {Zhang}}, \bibinfo
  {author} {\bibfnamefont {Y.}~\bibnamefont {Huang}}, \bibinfo {author}
  {\bibfnamefont {I.}~\bibnamefont {Pletikosi{\'{c}}}}, \bibinfo {author}
  {\bibfnamefont {A.}~\bibnamefont {Fedorov}}, \bibinfo {author} {\bibfnamefont
  {R.}~\bibnamefont {Zhong}}, \bibinfo {author} {\bibfnamefont
  {J.}~\bibnamefont {Schneeloch}}, \bibinfo {author} {\bibfnamefont
  {G.}~\bibnamefont {Gu}}, \ and\ \bibinfo {author} {\bibfnamefont
  {T.}~\bibnamefont {Valla}},\ }\href {\doibase 10.1038/nphys3648} {\bibfield
  {journal} {\bibinfo  {journal} {Nature Physics}\ }\textbf {\bibinfo {volume}
  {12}},\ \bibinfo {pages} {550} (\bibinfo {year}
  {2016}{\natexlab{b}})}\BibitemShut {NoStop}%
\bibitem [{\citenamefont {Hirschberger}\ \emph {et~al.}(2016)\citenamefont
  {Hirschberger}, \citenamefont {Kushwaha}, \citenamefont {Wang}, \citenamefont
  {Gibson}, \citenamefont {Liang}, \citenamefont {Belvin}, \citenamefont
  {Bernevig}, \citenamefont {Cava},\ and\ \citenamefont {Ong}}]{PHEGdPtBi}%
  \BibitemOpen
  \bibfield  {author} {\bibinfo {author} {\bibfnamefont {M.}~\bibnamefont
  {Hirschberger}}, \bibinfo {author} {\bibfnamefont {S.}~\bibnamefont
  {Kushwaha}}, \bibinfo {author} {\bibfnamefont {Z.}~\bibnamefont {Wang}},
  \bibinfo {author} {\bibfnamefont {Q.}~\bibnamefont {Gibson}}, \bibinfo
  {author} {\bibfnamefont {S.}~\bibnamefont {Liang}}, \bibinfo {author}
  {\bibfnamefont {C.}~\bibnamefont {Belvin}}, \bibinfo {author} {\bibfnamefont
  {B.}~\bibnamefont {Bernevig}}, \bibinfo {author} {\bibfnamefont
  {R.}~\bibnamefont {Cava}}, \ and\ \bibinfo {author} {\bibfnamefont
  {N.}~\bibnamefont {Ong}},\ }\href {\doibase 10.1038/nmat4684} {\bibfield
  {journal} {\bibinfo  {journal} {Nature Materials}\ }\textbf {\bibinfo
  {volume} {15}},\ \bibinfo {pages} {1161} (\bibinfo {year}
  {2016})}\BibitemShut {NoStop}%
\bibitem [{\citenamefont {Kumar}\ \emph {et~al.}(2018)\citenamefont {Kumar},
  \citenamefont {Guin}, \citenamefont {Felser},\ and\ \citenamefont
  {Shekhar}}]{PHEGdPtBiprb}%
  \BibitemOpen
  \bibfield  {author} {\bibinfo {author} {\bibfnamefont {N.}~\bibnamefont
  {Kumar}}, \bibinfo {author} {\bibfnamefont {S.~N.}\ \bibnamefont {Guin}},
  \bibinfo {author} {\bibfnamefont {C.}~\bibnamefont {Felser}}, \ and\ \bibinfo
  {author} {\bibfnamefont {C.}~\bibnamefont {Shekhar}},\ }\href {\doibase
  10.1103/PhysRevB.98.041103} {\bibfield  {journal} {\bibinfo  {journal} {Phys.
  Rev. B}\ }\textbf {\bibinfo {volume} {98}},\ \bibinfo {pages} {041103}
  (\bibinfo {year} {2018})}\BibitemShut {NoStop}%
\bibitem [{\citenamefont {Taskin}\ \emph {et~al.}(2017)\citenamefont {Taskin},
  \citenamefont {Legg}, \citenamefont {Yang}, \citenamefont {Sasaki},
  \citenamefont {Kanai}, \citenamefont {Matsumoto}, \citenamefont {Rosch},\
  and\ \citenamefont {Ando}}]{PHEBiSbTe}%
  \BibitemOpen
  \bibfield  {author} {\bibinfo {author} {\bibfnamefont {A.~A.}\ \bibnamefont
  {Taskin}}, \bibinfo {author} {\bibfnamefont {H.~F.}\ \bibnamefont {Legg}},
  \bibinfo {author} {\bibfnamefont {F.}~\bibnamefont {Yang}}, \bibinfo {author}
  {\bibfnamefont {S.}~\bibnamefont {Sasaki}}, \bibinfo {author} {\bibfnamefont
  {Y.}~\bibnamefont {Kanai}}, \bibinfo {author} {\bibfnamefont
  {K.}~\bibnamefont {Matsumoto}}, \bibinfo {author} {\bibfnamefont
  {A.}~\bibnamefont {Rosch}}, \ and\ \bibinfo {author} {\bibfnamefont
  {Y.}~\bibnamefont {Ando}},\ }\href {\doibase 10.1038/s41467-017-01474-8}
  {\bibfield  {journal} {\bibinfo  {journal} {Nature Communications}\ }\textbf
  {\bibinfo {volume} {8}},\ \bibinfo {pages} {1340} (\bibinfo {year}
  {2017})}\BibitemShut {NoStop}%
\bibitem [{\citenamefont {Li}\ \emph {et~al.}(2018{\natexlab{a}})\citenamefont
  {Li}, \citenamefont {Wang}, \citenamefont {He}, \citenamefont {Wang},\ and\
  \citenamefont {Shen}}]{PHECd3As2}%
  \BibitemOpen
  \bibfield  {author} {\bibinfo {author} {\bibfnamefont {H.}~\bibnamefont
  {Li}}, \bibinfo {author} {\bibfnamefont {H.-W.}\ \bibnamefont {Wang}},
  \bibinfo {author} {\bibfnamefont {H.}~\bibnamefont {He}}, \bibinfo {author}
  {\bibfnamefont {J.}~\bibnamefont {Wang}}, \ and\ \bibinfo {author}
  {\bibfnamefont {S.-Q.}\ \bibnamefont {Shen}},\ }\href {\doibase
  10.1103/PhysRevB.97.201110} {\bibfield  {journal} {\bibinfo  {journal} {Phys.
  Rev. B}\ }\textbf {\bibinfo {volume} {97}},\ \bibinfo {pages} {201110}
  (\bibinfo {year} {2018}{\natexlab{a}})}\BibitemShut {NoStop}%
\bibitem [{\citenamefont {Li}\ \emph {et~al.}(2018{\natexlab{b}})\citenamefont
  {Li}, \citenamefont {Zhang}, \citenamefont {Zhang}, \citenamefont {Wen},\
  and\ \citenamefont {Zhang}}]{PHEZrTe5prb}%
  \BibitemOpen
  \bibfield  {author} {\bibinfo {author} {\bibfnamefont {P.}~\bibnamefont
  {Li}}, \bibinfo {author} {\bibfnamefont {C.~H.}\ \bibnamefont {Zhang}},
  \bibinfo {author} {\bibfnamefont {J.~W.}\ \bibnamefont {Zhang}}, \bibinfo
  {author} {\bibfnamefont {Y.}~\bibnamefont {Wen}}, \ and\ \bibinfo {author}
  {\bibfnamefont {X.~X.}\ \bibnamefont {Zhang}},\ }\href {\doibase
  10.1103/PhysRevB.98.121108} {\bibfield  {journal} {\bibinfo  {journal} {Phys.
  Rev. B}\ }\textbf {\bibinfo {volume} {98}},\ \bibinfo {pages} {121108}
  (\bibinfo {year} {2018}{\natexlab{b}})}\BibitemShut {NoStop}%
\bibitem [{\citenamefont {Li}\ \emph {et~al.}(2019)\citenamefont {Li},
  \citenamefont {Zhang}, \citenamefont {Wen}, \citenamefont {Cheng},
  \citenamefont {Nichols}, \citenamefont {Cory}, \citenamefont {Miao},\ and\
  \citenamefont {Zhang}}]{PHEWTe2}%
  \BibitemOpen
  \bibfield  {author} {\bibinfo {author} {\bibfnamefont {P.}~\bibnamefont
  {Li}}, \bibinfo {author} {\bibfnamefont {C.}~\bibnamefont {Zhang}}, \bibinfo
  {author} {\bibfnamefont {Y.}~\bibnamefont {Wen}}, \bibinfo {author}
  {\bibfnamefont {L.}~\bibnamefont {Cheng}}, \bibinfo {author} {\bibfnamefont
  {G.}~\bibnamefont {Nichols}}, \bibinfo {author} {\bibfnamefont {D.~G.}\
  \bibnamefont {Cory}}, \bibinfo {author} {\bibfnamefont {G.-X.}\ \bibnamefont
  {Miao}}, \ and\ \bibinfo {author} {\bibfnamefont {X.-X.}\ \bibnamefont
  {Zhang}},\ }\href {\doibase 10.1103/PhysRevB.100.205128} {\bibfield
  {journal} {\bibinfo  {journal} {Phys. Rev. B}\ }\textbf {\bibinfo {volume}
  {100}},\ \bibinfo {pages} {205128} (\bibinfo {year} {2019})}\BibitemShut
  {NoStop}%
\bibitem [{\citenamefont {Zhou}\ \emph {et~al.}(2019)\citenamefont {Zhou},
  \citenamefont {Ye}, \citenamefont {Gan}, \citenamefont {Tang}, \citenamefont
  {Chen}, \citenamefont {Du}, \citenamefont {Tian}, \citenamefont {Deng},
  \citenamefont {Guo}, \citenamefont {Lu}, \citenamefont {Liu},\ and\
  \citenamefont {He}}]{PHESmB6}%
  \BibitemOpen
  \bibfield  {author} {\bibinfo {author} {\bibfnamefont {L.}~\bibnamefont
  {Zhou}}, \bibinfo {author} {\bibfnamefont {B.~C.}\ \bibnamefont {Ye}},
  \bibinfo {author} {\bibfnamefont {H.~B.}\ \bibnamefont {Gan}}, \bibinfo
  {author} {\bibfnamefont {J.~Y.}\ \bibnamefont {Tang}}, \bibinfo {author}
  {\bibfnamefont {P.~B.}\ \bibnamefont {Chen}}, \bibinfo {author}
  {\bibfnamefont {Z.~Z.}\ \bibnamefont {Du}}, \bibinfo {author} {\bibfnamefont
  {Y.}~\bibnamefont {Tian}}, \bibinfo {author} {\bibfnamefont {S.~Z.}\
  \bibnamefont {Deng}}, \bibinfo {author} {\bibfnamefont {G.~P.}\ \bibnamefont
  {Guo}}, \bibinfo {author} {\bibfnamefont {H.~Z.}\ \bibnamefont {Lu}},
  \bibinfo {author} {\bibfnamefont {F.}~\bibnamefont {Liu}}, \ and\ \bibinfo
  {author} {\bibfnamefont {H.~T.}\ \bibnamefont {He}},\ }\href {\doibase
  10.1103/PhysRevB.99.155424} {\bibfield  {journal} {\bibinfo  {journal} {Phys.
  Rev. B}\ }\textbf {\bibinfo {volume} {99}},\ \bibinfo {pages} {155424}
  (\bibinfo {year} {2019})}\BibitemShut {NoStop}%
\bibitem [{\citenamefont {Zhang}\ \emph {et~al.}(2020)\citenamefont {Zhang},
  \citenamefont {Zhao}, \citenamefont {Li}, \citenamefont {Wang}, \citenamefont
  {Xie}, \citenamefont {Cheng}, \citenamefont {Li}, \citenamefont {Lin},
  \citenamefont {Xi}, \citenamefont {Ke}, \citenamefont {Yang}, \citenamefont
  {He}, \citenamefont {Sun}, \citenamefont {Wang}, \citenamefont {Zhang},\ and\
  \citenamefont {Zeng}}]{PHETe}%
  \BibitemOpen
  \bibfield  {author} {\bibinfo {author} {\bibfnamefont {N.}~\bibnamefont
  {Zhang}}, \bibinfo {author} {\bibfnamefont {G.}~\bibnamefont {Zhao}},
  \bibinfo {author} {\bibfnamefont {L.}~\bibnamefont {Li}}, \bibinfo {author}
  {\bibfnamefont {P.}~\bibnamefont {Wang}}, \bibinfo {author} {\bibfnamefont
  {L.}~\bibnamefont {Xie}}, \bibinfo {author} {\bibfnamefont {B.}~\bibnamefont
  {Cheng}}, \bibinfo {author} {\bibfnamefont {H.}~\bibnamefont {Li}}, \bibinfo
  {author} {\bibfnamefont {Z.}~\bibnamefont {Lin}}, \bibinfo {author}
  {\bibfnamefont {C.}~\bibnamefont {Xi}}, \bibinfo {author} {\bibfnamefont
  {J.}~\bibnamefont {Ke}}, \bibinfo {author} {\bibfnamefont {M.}~\bibnamefont
  {Yang}}, \bibinfo {author} {\bibfnamefont {J.}~\bibnamefont {He}}, \bibinfo
  {author} {\bibfnamefont {Z.}~\bibnamefont {Sun}}, \bibinfo {author}
  {\bibfnamefont {Z.}~\bibnamefont {Wang}}, \bibinfo {author} {\bibfnamefont
  {Z.}~\bibnamefont {Zhang}}, \ and\ \bibinfo {author} {\bibfnamefont
  {C.}~\bibnamefont {Zeng}},\ }\href {\doibase 10.1073/pnas.2002913117}
  {\bibfield  {journal} {\bibinfo  {journal} {Proceedings of the National
  Academy of Sciences}\ }\textbf {\bibinfo {volume} {117}},\ \bibinfo {pages}
  {11337} (\bibinfo {year} {2020})}\BibitemShut {NoStop}%
\bibitem [{\citenamefont {Huang}\ \emph {et~al.}(2021)\citenamefont {Huang},
  \citenamefont {Nakamura},\ and\ \citenamefont {Takagi}}]{PHESr3SnO}%
  \BibitemOpen
  \bibfield  {author} {\bibinfo {author} {\bibfnamefont {D.}~\bibnamefont
  {Huang}}, \bibinfo {author} {\bibfnamefont {H.}~\bibnamefont {Nakamura}}, \
  and\ \bibinfo {author} {\bibfnamefont {H.}~\bibnamefont {Takagi}},\ }\href
  {\doibase 10.1103/PhysRevResearch.3.013268} {\bibfield  {journal} {\bibinfo
  {journal} {Phys. Rev. Res.}\ }\textbf {\bibinfo {volume} {3}},\ \bibinfo
  {pages} {013268} (\bibinfo {year} {2021})}\BibitemShut {NoStop}%
\bibitem [{\citenamefont {Wang}\ \emph {et~al.}(2022)\citenamefont {Wang},
  \citenamefont {Huang}, \citenamefont {Liu}, \citenamefont {Feng},
  \citenamefont {Zhu}, \citenamefont {Wu}, \citenamefont {Xiao},\ and\
  \citenamefont {Yang}}]{PHEyang2022}%
  \BibitemOpen
  \bibfield  {author} {\bibinfo {author} {\bibfnamefont {H.}~\bibnamefont
  {Wang}}, \bibinfo {author} {\bibfnamefont {Y.-X.}\ \bibnamefont {Huang}},
  \bibinfo {author} {\bibfnamefont {H.}~\bibnamefont {Liu}}, \bibinfo {author}
  {\bibfnamefont {X.}~\bibnamefont {Feng}}, \bibinfo {author} {\bibfnamefont
  {J.}~\bibnamefont {Zhu}}, \bibinfo {author} {\bibfnamefont {W.}~\bibnamefont
  {Wu}}, \bibinfo {author} {\bibfnamefont {C.}~\bibnamefont {Xiao}}, \ and\
  \bibinfo {author} {\bibfnamefont {S.~A.}\ \bibnamefont {Yang}},\ }\href@noop
  {} {\enquote {\bibinfo {title} {Theory of intrinsic in-plane hall effect},}\
  } (\bibinfo {year} {2022}),\ \Eprint {http://arxiv.org/abs/2211.05978}
  {arXiv:2211.05978 [cond-mat.mes-hall]} \BibitemShut {NoStop}%
\bibitem [{\citenamefont {Zhou}\ \emph {et~al.}(2022)\citenamefont {Zhou},
  \citenamefont {Zhang}, \citenamefont {Lin}, \citenamefont {Cao},
  \citenamefont {Zhou}, \citenamefont {Jiang}, \citenamefont {Du},
  \citenamefont {Tang}, \citenamefont {Shi}, \citenamefont {Jiang},
  \citenamefont {Cao}, \citenamefont {Lin}, \citenamefont {Fu}, \citenamefont
  {Zhu}, \citenamefont {Guo}, \citenamefont {Huang}, \citenamefont {Yao},
  \citenamefont {Parkin}, \citenamefont {Zhou}, \citenamefont {Gao},
  \citenamefont {Wang}, \citenamefont {Hou}, \citenamefont {Yao}, \citenamefont
  {Suenaga}, \citenamefont {Wu},\ and\ \citenamefont {Liu}}]{InplaneNature}%
  \BibitemOpen
  \bibfield  {author} {\bibinfo {author} {\bibfnamefont {J.}~\bibnamefont
  {Zhou}}, \bibinfo {author} {\bibfnamefont {W.}~\bibnamefont {Zhang}},
  \bibinfo {author} {\bibfnamefont {Y.-C.}\ \bibnamefont {Lin}}, \bibinfo
  {author} {\bibfnamefont {J.}~\bibnamefont {Cao}}, \bibinfo {author}
  {\bibfnamefont {Y.}~\bibnamefont {Zhou}}, \bibinfo {author} {\bibfnamefont
  {W.}~\bibnamefont {Jiang}}, \bibinfo {author} {\bibfnamefont
  {H.}~\bibnamefont {Du}}, \bibinfo {author} {\bibfnamefont {B.}~\bibnamefont
  {Tang}}, \bibinfo {author} {\bibfnamefont {J.}~\bibnamefont {Shi}}, \bibinfo
  {author} {\bibfnamefont {B.}~\bibnamefont {Jiang}}, \bibinfo {author}
  {\bibfnamefont {X.}~\bibnamefont {Cao}}, \bibinfo {author} {\bibfnamefont
  {B.}~\bibnamefont {Lin}}, \bibinfo {author} {\bibfnamefont {Q.}~\bibnamefont
  {Fu}}, \bibinfo {author} {\bibfnamefont {C.}~\bibnamefont {Zhu}}, \bibinfo
  {author} {\bibfnamefont {W.}~\bibnamefont {Guo}}, \bibinfo {author}
  {\bibfnamefont {Y.}~\bibnamefont {Huang}}, \bibinfo {author} {\bibfnamefont
  {Y.}~\bibnamefont {Yao}}, \bibinfo {author} {\bibfnamefont {S.~S.~P.}\
  \bibnamefont {Parkin}}, \bibinfo {author} {\bibfnamefont {J.}~\bibnamefont
  {Zhou}}, \bibinfo {author} {\bibfnamefont {Y.}~\bibnamefont {Gao}}, \bibinfo
  {author} {\bibfnamefont {Y.}~\bibnamefont {Wang}}, \bibinfo {author}
  {\bibfnamefont {Y.}~\bibnamefont {Hou}}, \bibinfo {author} {\bibfnamefont
  {Y.}~\bibnamefont {Yao}}, \bibinfo {author} {\bibfnamefont {K.}~\bibnamefont
  {Suenaga}}, \bibinfo {author} {\bibfnamefont {X.}~\bibnamefont {Wu}}, \ and\
  \bibinfo {author} {\bibfnamefont {Z.}~\bibnamefont {Liu}},\ }\href {\doibase
  10.1038/s41586-022-05031-2} {\bibfield  {journal} {\bibinfo  {journal}
  {Nature}\ }\textbf {\bibinfo {volume} {609}},\ \bibinfo {pages} {46}
  (\bibinfo {year} {2022})}\BibitemShut {NoStop}%
\bibitem [{\citenamefont {Kohler}(1938)}]{Kohlerrule}%
  \BibitemOpen
  \bibfield  {author} {\bibinfo {author} {\bibfnamefont {M.}~\bibnamefont
  {Kohler}},\ }\href {\doibase https://doi.org/10.1002/andp.19384240124}
  {\bibfield  {journal} {\bibinfo  {journal} {Annalen der Physik}\ }\textbf
  {\bibinfo {volume} {424}},\ \bibinfo {pages} {211} (\bibinfo {year}
  {1938})}\BibitemShut {NoStop}%
\bibitem [{\citenamefont {Kohler}(1949)}]{Kohlerfig}%
  \BibitemOpen
  \bibfield  {author} {\bibinfo {author} {\bibfnamefont {M.}~\bibnamefont
  {Kohler}},\ }\href {\doibase 10.1007/BF00626581} {\bibfield  {journal}
  {\bibinfo  {journal} {Naturwissenschaften}\ }\textbf {\bibinfo {volume}
  {36}},\ \bibinfo {pages} {186} (\bibinfo {year} {1949})}\BibitemShut
  {NoStop}%
\bibitem [{\citenamefont {Ali}\ \emph {et~al.}(2014)\citenamefont {Ali},
  \citenamefont {Xiong}, \citenamefont {Flynn}, \citenamefont {Tao},
  \citenamefont {Gibson}, \citenamefont {Schoop}, \citenamefont {Liang},
  \citenamefont {Haldolaarachchige}, \citenamefont {Hirschberger},
  \citenamefont {Ong},\ and\ \citenamefont {Cava}}]{MRWTe2nature}%
  \BibitemOpen
  \bibfield  {author} {\bibinfo {author} {\bibfnamefont {M.~N.}\ \bibnamefont
  {Ali}}, \bibinfo {author} {\bibfnamefont {J.}~\bibnamefont {Xiong}}, \bibinfo
  {author} {\bibfnamefont {S.}~\bibnamefont {Flynn}}, \bibinfo {author}
  {\bibfnamefont {J.}~\bibnamefont {Tao}}, \bibinfo {author} {\bibfnamefont
  {Q.~D.}\ \bibnamefont {Gibson}}, \bibinfo {author} {\bibfnamefont {L.~M.}\
  \bibnamefont {Schoop}}, \bibinfo {author} {\bibfnamefont {T.}~\bibnamefont
  {Liang}}, \bibinfo {author} {\bibfnamefont {N.}~\bibnamefont
  {Haldolaarachchige}}, \bibinfo {author} {\bibfnamefont {M.}~\bibnamefont
  {Hirschberger}}, \bibinfo {author} {\bibfnamefont {N.~P.}\ \bibnamefont
  {Ong}}, \ and\ \bibinfo {author} {\bibfnamefont {R.~J.}\ \bibnamefont
  {Cava}},\ }\href {\doibase 10.1038/nature13763} {\bibfield  {journal}
  {\bibinfo  {journal} {Nature}\ }\textbf {\bibinfo {volume} {514}},\ \bibinfo
  {pages} {205} (\bibinfo {year} {2014})}\BibitemShut {NoStop}%
\bibitem [{\citenamefont {Wang}\ \emph {et~al.}(2015)\citenamefont {Wang},
  \citenamefont {Thoutam}, \citenamefont {Xiao}, \citenamefont {Hu},
  \citenamefont {Das}, \citenamefont {Mao}, \citenamefont {Wei}, \citenamefont
  {Divan}, \citenamefont {Luican-Mayer}, \citenamefont {Crabtree},\ and\
  \citenamefont {Kwok}}]{turnonKohler}%
  \BibitemOpen
  \bibfield  {author} {\bibinfo {author} {\bibfnamefont {Y.~L.}\ \bibnamefont
  {Wang}}, \bibinfo {author} {\bibfnamefont {L.~R.}\ \bibnamefont {Thoutam}},
  \bibinfo {author} {\bibfnamefont {Z.~L.}\ \bibnamefont {Xiao}}, \bibinfo
  {author} {\bibfnamefont {J.}~\bibnamefont {Hu}}, \bibinfo {author}
  {\bibfnamefont {S.}~\bibnamefont {Das}}, \bibinfo {author} {\bibfnamefont
  {Z.~Q.}\ \bibnamefont {Mao}}, \bibinfo {author} {\bibfnamefont
  {J.}~\bibnamefont {Wei}}, \bibinfo {author} {\bibfnamefont {R.}~\bibnamefont
  {Divan}}, \bibinfo {author} {\bibfnamefont {A.}~\bibnamefont {Luican-Mayer}},
  \bibinfo {author} {\bibfnamefont {G.~W.}\ \bibnamefont {Crabtree}}, \ and\
  \bibinfo {author} {\bibfnamefont {W.~K.}\ \bibnamefont {Kwok}},\ }\href
  {\doibase 10.1103/PhysRevB.92.180402} {\bibfield  {journal} {\bibinfo
  {journal} {Phys. Rev. B}\ }\textbf {\bibinfo {volume} {92}},\ \bibinfo
  {pages} {180402} (\bibinfo {year} {2015})}\BibitemShut {NoStop}%
\bibitem [{\citenamefont {Han}\ \emph {et~al.}(2017)\citenamefont {Han},
  \citenamefont {Xu}, \citenamefont {Botana}, \citenamefont {Xiao},
  \citenamefont {Wang}, \citenamefont {Yang}, \citenamefont {Chung},
  \citenamefont {Kanatzidis}, \citenamefont {Norman}, \citenamefont
  {Crabtree},\ and\ \citenamefont {Kwok}}]{turnonLasb}%
  \BibitemOpen
  \bibfield  {author} {\bibinfo {author} {\bibfnamefont {F.}~\bibnamefont
  {Han}}, \bibinfo {author} {\bibfnamefont {J.}~\bibnamefont {Xu}}, \bibinfo
  {author} {\bibfnamefont {A.~S.}\ \bibnamefont {Botana}}, \bibinfo {author}
  {\bibfnamefont {Z.~L.}\ \bibnamefont {Xiao}}, \bibinfo {author}
  {\bibfnamefont {Y.~L.}\ \bibnamefont {Wang}}, \bibinfo {author}
  {\bibfnamefont {W.~G.}\ \bibnamefont {Yang}}, \bibinfo {author}
  {\bibfnamefont {D.~Y.}\ \bibnamefont {Chung}}, \bibinfo {author}
  {\bibfnamefont {M.~G.}\ \bibnamefont {Kanatzidis}}, \bibinfo {author}
  {\bibfnamefont {M.~R.}\ \bibnamefont {Norman}}, \bibinfo {author}
  {\bibfnamefont {G.~W.}\ \bibnamefont {Crabtree}}, \ and\ \bibinfo {author}
  {\bibfnamefont {W.~K.}\ \bibnamefont {Kwok}},\ }\href {\doibase
  10.1103/PhysRevB.96.125112} {\bibfield  {journal} {\bibinfo  {journal} {Phys.
  Rev. B}\ }\textbf {\bibinfo {volume} {96}},\ \bibinfo {pages} {125112}
  (\bibinfo {year} {2017})}\BibitemShut {NoStop}%
\bibitem [{\citenamefont {Du}\ \emph {et~al.}(2018)\citenamefont {Du},
  \citenamefont {Lou}, \citenamefont {Zhang}, \citenamefont {Zhou},
  \citenamefont {Xu}, \citenamefont {Chen}, \citenamefont {Tang}, \citenamefont
  {Chen}, \citenamefont {Chen}, \citenamefont {Zhu}, \citenamefont {Wang},
  \citenamefont {Yang}, \citenamefont {Wu}, \citenamefont {Yazyev},\ and\
  \citenamefont {Fang}}]{MRalphaWP2}%
  \BibitemOpen
  \bibfield  {author} {\bibinfo {author} {\bibfnamefont {J.}~\bibnamefont
  {Du}}, \bibinfo {author} {\bibfnamefont {Z.}~\bibnamefont {Lou}}, \bibinfo
  {author} {\bibfnamefont {S.}~\bibnamefont {Zhang}}, \bibinfo {author}
  {\bibfnamefont {Y.}~\bibnamefont {Zhou}}, \bibinfo {author} {\bibfnamefont
  {B.}~\bibnamefont {Xu}}, \bibinfo {author} {\bibfnamefont {Q.}~\bibnamefont
  {Chen}}, \bibinfo {author} {\bibfnamefont {Y.}~\bibnamefont {Tang}}, \bibinfo
  {author} {\bibfnamefont {S.}~\bibnamefont {Chen}}, \bibinfo {author}
  {\bibfnamefont {H.}~\bibnamefont {Chen}}, \bibinfo {author} {\bibfnamefont
  {Q.}~\bibnamefont {Zhu}}, \bibinfo {author} {\bibfnamefont {H.}~\bibnamefont
  {Wang}}, \bibinfo {author} {\bibfnamefont {J.}~\bibnamefont {Yang}}, \bibinfo
  {author} {\bibfnamefont {Q.}~\bibnamefont {Wu}}, \bibinfo {author}
  {\bibfnamefont {O.~V.}\ \bibnamefont {Yazyev}}, \ and\ \bibinfo {author}
  {\bibfnamefont {M.}~\bibnamefont {Fang}},\ }\href {\doibase
  10.1103/PhysRevB.97.245101} {\bibfield  {journal} {\bibinfo  {journal} {Phys.
  Rev. B}\ }\textbf {\bibinfo {volume} {97}},\ \bibinfo {pages} {245101}
  (\bibinfo {year} {2018})}\BibitemShut {NoStop}%
\bibitem [{\citenamefont {Pavlosiuk}\ \emph {et~al.}(2018)\citenamefont
  {Pavlosiuk}, \citenamefont {Swatek}, \citenamefont {Kaczorowski},\ and\
  \citenamefont {Wi\ifmmode~\acute{s}\else \'{s}\fi{}niewski}}]{turnonYBi}%
  \BibitemOpen
  \bibfield  {author} {\bibinfo {author} {\bibfnamefont {O.}~\bibnamefont
  {Pavlosiuk}}, \bibinfo {author} {\bibfnamefont {P.}~\bibnamefont {Swatek}},
  \bibinfo {author} {\bibfnamefont {D.}~\bibnamefont {Kaczorowski}}, \ and\
  \bibinfo {author} {\bibfnamefont {P.}~\bibnamefont {Wi\ifmmode~\acute{s}\else
  \'{s}\fi{}niewski}},\ }\href {\doibase 10.1103/PhysRevB.97.235132} {\bibfield
   {journal} {\bibinfo  {journal} {Phys. Rev. B}\ }\textbf {\bibinfo {volume}
  {97}},\ \bibinfo {pages} {235132} (\bibinfo {year} {2018})}\BibitemShut
  {NoStop}%
\bibitem [{\citenamefont {Gatti}\ \emph {et~al.}(2021)\citenamefont {Gatti},
  \citenamefont {Gos\'albez-Mart\'{\i}nez}, \citenamefont {Wu}, \citenamefont
  {Hu}, \citenamefont {Zhang}, \citenamefont {Aut\`es}, \citenamefont {Puppin},
  \citenamefont {Bugini}, \citenamefont {Berger}, \citenamefont {Moreschini},
  \citenamefont {Vobornik}, \citenamefont {Fujii}, \citenamefont {Ansermet},
  \citenamefont {Yazyev},\ and\ \citenamefont {Crepaldi}}]{MRTaSe3}%
  \BibitemOpen
  \bibfield  {author} {\bibinfo {author} {\bibfnamefont {G.}~\bibnamefont
  {Gatti}}, \bibinfo {author} {\bibfnamefont {D.}~\bibnamefont
  {Gos\'albez-Mart\'{\i}nez}}, \bibinfo {author} {\bibfnamefont {Q.~S.}\
  \bibnamefont {Wu}}, \bibinfo {author} {\bibfnamefont {J.}~\bibnamefont {Hu}},
  \bibinfo {author} {\bibfnamefont {S.~N.}\ \bibnamefont {Zhang}}, \bibinfo
  {author} {\bibfnamefont {G.}~\bibnamefont {Aut\`es}}, \bibinfo {author}
  {\bibfnamefont {M.}~\bibnamefont {Puppin}}, \bibinfo {author} {\bibfnamefont
  {D.}~\bibnamefont {Bugini}}, \bibinfo {author} {\bibfnamefont
  {H.}~\bibnamefont {Berger}}, \bibinfo {author} {\bibfnamefont
  {L.}~\bibnamefont {Moreschini}}, \bibinfo {author} {\bibfnamefont
  {I.}~\bibnamefont {Vobornik}}, \bibinfo {author} {\bibfnamefont
  {J.}~\bibnamefont {Fujii}}, \bibinfo {author} {\bibfnamefont {J.-P.}\
  \bibnamefont {Ansermet}}, \bibinfo {author} {\bibfnamefont {O.~V.}\
  \bibnamefont {Yazyev}}, \ and\ \bibinfo {author} {\bibfnamefont
  {A.}~\bibnamefont {Crepaldi}},\ }\href {\doibase 10.1103/PhysRevB.104.155122}
  {\bibfield  {journal} {\bibinfo  {journal} {Phys. Rev. B}\ }\textbf {\bibinfo
  {volume} {104}},\ \bibinfo {pages} {155122} (\bibinfo {year}
  {2021})}\BibitemShut {NoStop}%
\bibitem [{\citenamefont {Chen}\ \emph {et~al.}(2020)\citenamefont {Chen},
  \citenamefont {Lou}, \citenamefont {Zhang}, \citenamefont {Xu}, \citenamefont
  {Zhou}, \citenamefont {Chen}, \citenamefont {Chen}, \citenamefont {Du},
  \citenamefont {Wang}, \citenamefont {Yang}, \citenamefont {Wu}, \citenamefont
  {Yazyev},\ and\ \citenamefont {Fang}}]{MRMoO2}%
  \BibitemOpen
  \bibfield  {author} {\bibinfo {author} {\bibfnamefont {Q.}~\bibnamefont
  {Chen}}, \bibinfo {author} {\bibfnamefont {Z.}~\bibnamefont {Lou}}, \bibinfo
  {author} {\bibfnamefont {S.}~\bibnamefont {Zhang}}, \bibinfo {author}
  {\bibfnamefont {B.}~\bibnamefont {Xu}}, \bibinfo {author} {\bibfnamefont
  {Y.}~\bibnamefont {Zhou}}, \bibinfo {author} {\bibfnamefont {H.}~\bibnamefont
  {Chen}}, \bibinfo {author} {\bibfnamefont {S.}~\bibnamefont {Chen}}, \bibinfo
  {author} {\bibfnamefont {J.}~\bibnamefont {Du}}, \bibinfo {author}
  {\bibfnamefont {H.}~\bibnamefont {Wang}}, \bibinfo {author} {\bibfnamefont
  {J.}~\bibnamefont {Yang}}, \bibinfo {author} {\bibfnamefont {Q.}~\bibnamefont
  {Wu}}, \bibinfo {author} {\bibfnamefont {O.~V.}\ \bibnamefont {Yazyev}}, \
  and\ \bibinfo {author} {\bibfnamefont {M.}~\bibnamefont {Fang}},\ }\href
  {\doibase 10.1103/PhysRevB.102.165133} {\bibfield  {journal} {\bibinfo
  {journal} {Phys. Rev. B}\ }\textbf {\bibinfo {volume} {102}},\ \bibinfo
  {pages} {165133} (\bibinfo {year} {2020})}\BibitemShut {NoStop}%
\bibitem [{\citenamefont {Chapai}\ \emph {et~al.}(2020)\citenamefont {Chapai},
  \citenamefont {Browne}, \citenamefont {Graf}, \citenamefont {DiTusa},\ and\
  \citenamefont {Jin}}]{turnonPdTe2}%
  \BibitemOpen
  \bibfield  {author} {\bibinfo {author} {\bibfnamefont {R.}~\bibnamefont
  {Chapai}}, \bibinfo {author} {\bibfnamefont {D.~A.}\ \bibnamefont {Browne}},
  \bibinfo {author} {\bibfnamefont {D.~E.}\ \bibnamefont {Graf}}, \bibinfo
  {author} {\bibfnamefont {J.~F.}\ \bibnamefont {DiTusa}}, \ and\ \bibinfo
  {author} {\bibfnamefont {R.}~\bibnamefont {Jin}},\ }\href {\doibase
  10.1088/1361-648X/abb548} {\bibfield  {journal} {\bibinfo  {journal} {Journal
  of Physics: Condensed Matter}\ }\textbf {\bibinfo {volume} {33}},\ \bibinfo
  {pages} {035601} (\bibinfo {year} {2020})}\BibitemShut {NoStop}%
\bibitem [{\citenamefont {Xu}\ \emph {et~al.}(2021)\citenamefont {Xu},
  \citenamefont {Han}, \citenamefont {Wang}, \citenamefont {Thoutam},
  \citenamefont {Pate}, \citenamefont {Li}, \citenamefont {Zhang},
  \citenamefont {Wang}, \citenamefont {Fotovat}, \citenamefont {Welp},
  \citenamefont {Zhou}, \citenamefont {Kwok}, \citenamefont {Chung},
  \citenamefont {Kanatzidis},\ and\ \citenamefont {Xiao}}]{ExtendedKohler}%
  \BibitemOpen
  \bibfield  {author} {\bibinfo {author} {\bibfnamefont {J.}~\bibnamefont
  {Xu}}, \bibinfo {author} {\bibfnamefont {F.}~\bibnamefont {Han}}, \bibinfo
  {author} {\bibfnamefont {T.-T.}\ \bibnamefont {Wang}}, \bibinfo {author}
  {\bibfnamefont {L.~R.}\ \bibnamefont {Thoutam}}, \bibinfo {author}
  {\bibfnamefont {S.~E.}\ \bibnamefont {Pate}}, \bibinfo {author}
  {\bibfnamefont {M.}~\bibnamefont {Li}}, \bibinfo {author} {\bibfnamefont
  {X.}~\bibnamefont {Zhang}}, \bibinfo {author} {\bibfnamefont {Y.-L.}\
  \bibnamefont {Wang}}, \bibinfo {author} {\bibfnamefont {R.}~\bibnamefont
  {Fotovat}}, \bibinfo {author} {\bibfnamefont {U.}~\bibnamefont {Welp}},
  \bibinfo {author} {\bibfnamefont {X.}~\bibnamefont {Zhou}}, \bibinfo {author}
  {\bibfnamefont {W.-K.}\ \bibnamefont {Kwok}}, \bibinfo {author}
  {\bibfnamefont {D.~Y.}\ \bibnamefont {Chung}}, \bibinfo {author}
  {\bibfnamefont {M.~G.}\ \bibnamefont {Kanatzidis}}, \ and\ \bibinfo {author}
  {\bibfnamefont {Z.-L.}\ \bibnamefont {Xiao}},\ }\href {\doibase
  10.1103/PhysRevX.11.041029} {\bibfield  {journal} {\bibinfo  {journal} {Phys.
  Rev. X}\ }\textbf {\bibinfo {volume} {11}},\ \bibinfo {pages} {041029}
  (\bibinfo {year} {2021})}\BibitemShut {NoStop}%
\bibitem [{\citenamefont {Zhang}\ \emph {et~al.}(2019)\citenamefont {Zhang},
  \citenamefont {Wu}, \citenamefont {Liu},\ and\ \citenamefont
  {Yazyev}}]{MRZhangprb}%
  \BibitemOpen
  \bibfield  {author} {\bibinfo {author} {\bibfnamefont {S.}~\bibnamefont
  {Zhang}}, \bibinfo {author} {\bibfnamefont {Q.}~\bibnamefont {Wu}}, \bibinfo
  {author} {\bibfnamefont {Y.}~\bibnamefont {Liu}}, \ and\ \bibinfo {author}
  {\bibfnamefont {O.~V.}\ \bibnamefont {Yazyev}},\ }\href {\doibase
  10.1103/PhysRevB.99.035142} {\bibfield  {journal} {\bibinfo  {journal} {Phys.
  Rev. B}\ }\textbf {\bibinfo {volume} {99}},\ \bibinfo {pages} {035142}
  (\bibinfo {year} {2019})}\BibitemShut {NoStop}%
\bibitem [{\citenamefont {Zhou}\ \emph {et~al.}(2020)\citenamefont {Zhou},
  \citenamefont {Lou}, \citenamefont {Zhang}, \citenamefont {Chen},
  \citenamefont {Chen}, \citenamefont {Xu}, \citenamefont {Du}, \citenamefont
  {Yang}, \citenamefont {Wang}, \citenamefont {Xi}, \citenamefont {Pi},
  \citenamefont {Wu}, \citenamefont {Yazyev},\ and\ \citenamefont
  {Fang}}]{MRSiP2}%
  \BibitemOpen
  \bibfield  {author} {\bibinfo {author} {\bibfnamefont {Y.}~\bibnamefont
  {Zhou}}, \bibinfo {author} {\bibfnamefont {Z.}~\bibnamefont {Lou}}, \bibinfo
  {author} {\bibfnamefont {S.}~\bibnamefont {Zhang}}, \bibinfo {author}
  {\bibfnamefont {H.}~\bibnamefont {Chen}}, \bibinfo {author} {\bibfnamefont
  {Q.}~\bibnamefont {Chen}}, \bibinfo {author} {\bibfnamefont {B.}~\bibnamefont
  {Xu}}, \bibinfo {author} {\bibfnamefont {J.}~\bibnamefont {Du}}, \bibinfo
  {author} {\bibfnamefont {J.}~\bibnamefont {Yang}}, \bibinfo {author}
  {\bibfnamefont {H.}~\bibnamefont {Wang}}, \bibinfo {author} {\bibfnamefont
  {C.}~\bibnamefont {Xi}}, \bibinfo {author} {\bibfnamefont {L.}~\bibnamefont
  {Pi}}, \bibinfo {author} {\bibfnamefont {Q.}~\bibnamefont {Wu}}, \bibinfo
  {author} {\bibfnamefont {O.~V.}\ \bibnamefont {Yazyev}}, \ and\ \bibinfo
  {author} {\bibfnamefont {M.}~\bibnamefont {Fang}},\ }\href {\doibase
  10.1103/PhysRevB.102.115145} {\bibfield  {journal} {\bibinfo  {journal}
  {Phys. Rev. B}\ }\textbf {\bibinfo {volume} {102}},\ \bibinfo {pages}
  {115145} (\bibinfo {year} {2020})}\BibitemShut {NoStop}%
\bibitem [{\citenamefont {Chen}\ \emph {et~al.}(2021)\citenamefont {Chen},
  \citenamefont {Lou}, \citenamefont {Zhang}, \citenamefont {Zhou},
  \citenamefont {Xu}, \citenamefont {Chen}, \citenamefont {Chen}, \citenamefont
  {Du}, \citenamefont {Wang}, \citenamefont {Yang}, \citenamefont {Wu},
  \citenamefont {Yazyev},\ and\ \citenamefont {Fang}}]{MRReO3}%
  \BibitemOpen
  \bibfield  {author} {\bibinfo {author} {\bibfnamefont {Q.}~\bibnamefont
  {Chen}}, \bibinfo {author} {\bibfnamefont {Z.}~\bibnamefont {Lou}}, \bibinfo
  {author} {\bibfnamefont {S.}~\bibnamefont {Zhang}}, \bibinfo {author}
  {\bibfnamefont {Y.}~\bibnamefont {Zhou}}, \bibinfo {author} {\bibfnamefont
  {B.}~\bibnamefont {Xu}}, \bibinfo {author} {\bibfnamefont {H.}~\bibnamefont
  {Chen}}, \bibinfo {author} {\bibfnamefont {S.}~\bibnamefont {Chen}}, \bibinfo
  {author} {\bibfnamefont {J.}~\bibnamefont {Du}}, \bibinfo {author}
  {\bibfnamefont {H.}~\bibnamefont {Wang}}, \bibinfo {author} {\bibfnamefont
  {J.}~\bibnamefont {Yang}}, \bibinfo {author} {\bibfnamefont {Q.}~\bibnamefont
  {Wu}}, \bibinfo {author} {\bibfnamefont {O.~V.}\ \bibnamefont {Yazyev}}, \
  and\ \bibinfo {author} {\bibfnamefont {M.}~\bibnamefont {Fang}},\ }\href
  {\doibase 10.1103/PhysRevB.104.115104} {\bibfield  {journal} {\bibinfo
  {journal} {Phys. Rev. B}\ }\textbf {\bibinfo {volume} {104}},\ \bibinfo
  {pages} {115104} (\bibinfo {year} {2021})}\BibitemShut {NoStop}%
\bibitem [{\citenamefont {Novak}\ \emph {et~al.}(2019)\citenamefont {Novak},
  \citenamefont {Zhang}, \citenamefont {Orbani\ifmmode~\acute{c}\else
  \'{c}\fi{}}, \citenamefont {Bili\ifmmode~\check{s}\else \v{s}\fi{}kov},
  \citenamefont {Eguchi}, \citenamefont {Paschen}, \citenamefont {Kimura},
  \citenamefont {Wang}, \citenamefont {Osada}, \citenamefont {Uchida},
  \citenamefont {Sato}, \citenamefont {Wu}, \citenamefont {Yazyev},\ and\
  \citenamefont {Kokanovi\ifmmode~\acute{c}\else \'{c}\fi{}}}]{MRZrSiS}%
  \BibitemOpen
  \bibfield  {author} {\bibinfo {author} {\bibfnamefont {M.}~\bibnamefont
  {Novak}}, \bibinfo {author} {\bibfnamefont {S.~N.}\ \bibnamefont {Zhang}},
  \bibinfo {author} {\bibfnamefont {F.}~\bibnamefont
  {Orbani\ifmmode~\acute{c}\else \'{c}\fi{}}}, \bibinfo {author} {\bibfnamefont
  {N.}~\bibnamefont {Bili\ifmmode~\check{s}\else \v{s}\fi{}kov}}, \bibinfo
  {author} {\bibfnamefont {G.}~\bibnamefont {Eguchi}}, \bibinfo {author}
  {\bibfnamefont {S.}~\bibnamefont {Paschen}}, \bibinfo {author} {\bibfnamefont
  {A.}~\bibnamefont {Kimura}}, \bibinfo {author} {\bibfnamefont {X.~X.}\
  \bibnamefont {Wang}}, \bibinfo {author} {\bibfnamefont {T.}~\bibnamefont
  {Osada}}, \bibinfo {author} {\bibfnamefont {K.}~\bibnamefont {Uchida}},
  \bibinfo {author} {\bibfnamefont {M.}~\bibnamefont {Sato}}, \bibinfo {author}
  {\bibfnamefont {Q.~S.}\ \bibnamefont {Wu}}, \bibinfo {author} {\bibfnamefont
  {O.~V.}\ \bibnamefont {Yazyev}}, \ and\ \bibinfo {author} {\bibfnamefont
  {I.}~\bibnamefont {Kokanovi\ifmmode~\acute{c}\else \'{c}\fi{}}},\ }\href
  {\doibase 10.1103/PhysRevB.100.085137} {\bibfield  {journal} {\bibinfo
  {journal} {Phys. Rev. B}\ }\textbf {\bibinfo {volume} {100}},\ \bibinfo
  {pages} {085137} (\bibinfo {year} {2019})}\BibitemShut {NoStop}%
\bibitem [{\citenamefont {Pi}\ \emph {et~al.}(2024)\citenamefont {Pi},
  \citenamefont {Zhang}, \citenamefont {Xu}, \citenamefont {Fang},
  \citenamefont {Weng},\ and\ \citenamefont {Wu}}]{pi2024_ZrTe5}%
  \BibitemOpen
  \bibfield  {author} {\bibinfo {author} {\bibfnamefont {H.}~\bibnamefont
  {Pi}}, \bibinfo {author} {\bibfnamefont {S.}~\bibnamefont {Zhang}}, \bibinfo
  {author} {\bibfnamefont {Y.}~\bibnamefont {Xu}}, \bibinfo {author}
  {\bibfnamefont {Z.}~\bibnamefont {Fang}}, \bibinfo {author} {\bibfnamefont
  {H.}~\bibnamefont {Weng}}, \ and\ \bibinfo {author} {\bibfnamefont
  {Q.}~\bibnamefont {Wu}},\ }\href {https://arxiv.org/abs/2401.15151} {\enquote
  {\bibinfo {title} {First-principles methodology for studying magnetotransport
  in narrow-gap semiconductors: an application to zirconium pentatelluride
  zrte5},}\ } (\bibinfo {year} {2024}),\ \Eprint
  {http://arxiv.org/abs/2401.15151} {arXiv:2401.15151 [cond-mat.mtrl-sci]}
  \BibitemShut {NoStop}%
\bibitem [{\citenamefont {Liu}\ \emph {et~al.}(2024{\natexlab{a}})\citenamefont
  {Liu}, \citenamefont {Zhang}, \citenamefont {Fang}, \citenamefont {Weng},\
  and\ \citenamefont {Wu}}]{zhliu2024}%
  \BibitemOpen
  \bibfield  {author} {\bibinfo {author} {\bibfnamefont {Z.}~\bibnamefont
  {Liu}}, \bibinfo {author} {\bibfnamefont {S.}~\bibnamefont {Zhang}}, \bibinfo
  {author} {\bibfnamefont {Z.}~\bibnamefont {Fang}}, \bibinfo {author}
  {\bibfnamefont {H.}~\bibnamefont {Weng}}, \ and\ \bibinfo {author}
  {\bibfnamefont {Q.}~\bibnamefont {Wu}},\ }\href
  {https://arxiv.org/abs/2401.15146} {\enquote {\bibinfo {title}
  {First-principles methodology for studying magnetotransport in magnetic
  materials},}\ } (\bibinfo {year} {2024}{\natexlab{a}}),\ \Eprint
  {http://arxiv.org/abs/2401.15146} {arXiv:2401.15146 [cond-mat.mtrl-sci]}
  \BibitemShut {NoStop}%
\bibitem [{\citenamefont {Kresse}\ and\ \citenamefont
  {Furthm\"uller}(1996)}]{Vasp1}%
  \BibitemOpen
  \bibfield  {author} {\bibinfo {author} {\bibfnamefont {G.}~\bibnamefont
  {Kresse}}\ and\ \bibinfo {author} {\bibfnamefont {J.}~\bibnamefont
  {Furthm\"uller}},\ }\href {\doibase 10.1103/PhysRevB.54.11169} {\bibfield
  {journal} {\bibinfo  {journal} {Phys. Rev. B}\ }\textbf {\bibinfo {volume}
  {54}},\ \bibinfo {pages} {11169} (\bibinfo {year} {1996})}\BibitemShut
  {NoStop}%
\bibitem [{\citenamefont {Kresse}\ and\ \citenamefont {Joubert}(1999)}]{Vasp2}%
  \BibitemOpen
  \bibfield  {author} {\bibinfo {author} {\bibfnamefont {G.}~\bibnamefont
  {Kresse}}\ and\ \bibinfo {author} {\bibfnamefont {D.}~\bibnamefont
  {Joubert}},\ }\href {\doibase 10.1103/PhysRevB.59.1758} {\bibfield  {journal}
  {\bibinfo  {journal} {Phys. Rev. B}\ }\textbf {\bibinfo {volume} {59}},\
  \bibinfo {pages} {1758} (\bibinfo {year} {1999})}\BibitemShut {NoStop}%
\bibitem [{\citenamefont {Mostofi}\ \emph {et~al.}(2014)\citenamefont
  {Mostofi}, \citenamefont {Yates}, \citenamefont {Pizzi}, \citenamefont {Lee},
  \citenamefont {Souza}, \citenamefont {Vanderbilt},\ and\ \citenamefont
  {Marzari}}]{NMwannier90}%
  \BibitemOpen
  \bibfield  {author} {\bibinfo {author} {\bibfnamefont {A.~A.}\ \bibnamefont
  {Mostofi}}, \bibinfo {author} {\bibfnamefont {J.~R.}\ \bibnamefont {Yates}},
  \bibinfo {author} {\bibfnamefont {G.}~\bibnamefont {Pizzi}}, \bibinfo
  {author} {\bibfnamefont {Y.-S.}\ \bibnamefont {Lee}}, \bibinfo {author}
  {\bibfnamefont {I.}~\bibnamefont {Souza}}, \bibinfo {author} {\bibfnamefont
  {D.}~\bibnamefont {Vanderbilt}}, \ and\ \bibinfo {author} {\bibfnamefont
  {N.}~\bibnamefont {Marzari}},\ }\href {\doibase
  https://doi.org/10.1016/j.cpc.2014.05.003} {\bibfield  {journal} {\bibinfo
  {journal} {Computer Physics Communications}\ }\textbf {\bibinfo {volume}
  {185}},\ \bibinfo {pages} {2309} (\bibinfo {year} {2014})}\BibitemShut
  {NoStop}%
\bibitem [{\citenamefont {Liu}\ \emph {et~al.}(2009)\citenamefont {Liu},
  \citenamefont {Zhang},\ and\ \citenamefont {Yao}}]{Liuyiprb}%
  \BibitemOpen
  \bibfield  {author} {\bibinfo {author} {\bibfnamefont {Y.}~\bibnamefont
  {Liu}}, \bibinfo {author} {\bibfnamefont {H.-J.}\ \bibnamefont {Zhang}}, \
  and\ \bibinfo {author} {\bibfnamefont {Y.}~\bibnamefont {Yao}},\ }\href
  {\doibase 10.1103/PhysRevB.79.245123} {\bibfield  {journal} {\bibinfo
  {journal} {Phys. Rev. B}\ }\textbf {\bibinfo {volume} {79}},\ \bibinfo
  {pages} {245123} (\bibinfo {year} {2009})}\BibitemShut {NoStop}%
\bibitem [{\citenamefont {Wu}\ \emph {et~al.}(2018)\citenamefont {Wu},
  \citenamefont {Zhang}, \citenamefont {Song}, \citenamefont {Troyer},\ and\
  \citenamefont {Soluyanov}}]{WUWT}%
  \BibitemOpen
  \bibfield  {author} {\bibinfo {author} {\bibfnamefont {Q.}~\bibnamefont
  {Wu}}, \bibinfo {author} {\bibfnamefont {S.}~\bibnamefont {Zhang}}, \bibinfo
  {author} {\bibfnamefont {H.-F.}\ \bibnamefont {Song}}, \bibinfo {author}
  {\bibfnamefont {M.}~\bibnamefont {Troyer}}, \ and\ \bibinfo {author}
  {\bibfnamefont {A.~A.}\ \bibnamefont {Soluyanov}},\ }\href {\doibase
  https://doi.org/10.1016/j.cpc.2017.09.033} {\bibfield  {journal} {\bibinfo
  {journal} {Computer Physics Communications}\ }\textbf {\bibinfo {volume}
  {224}},\ \bibinfo {pages} {405} (\bibinfo {year} {2018})}\BibitemShut
  {NoStop}%
\bibitem [{sup(2023)}]{supp}%
  \BibitemOpen
  \href@noop {} {}\bibinfo {howpublished}
  {\url{URL_will_be_inserted_by_publisher}} (\bibinfo {year} {2023}),\ \bibinfo
  {note} {supplementary materials}\BibitemShut {NoStop}%
\bibitem [{\citenamefont {Fawcett}(1964)}]{Fawcett1964}%
  \BibitemOpen
  \bibfield  {author} {\bibinfo {author} {\bibfnamefont {E.}~\bibnamefont
  {Fawcett}},\ }\href {\doibase 10.1080/00018736400101021} {\bibfield
  {journal} {\bibinfo  {journal} {Advances in Physics}\ }\textbf {\bibinfo
  {volume} {13}},\ \bibinfo {pages} {139–191} (\bibinfo {year}
  {1964})}\BibitemShut {NoStop}%
\bibitem [{\citenamefont {Pavlosiuk}\ and\ \citenamefont
  {Kaczorowski}(2018)}]{Ptte2nasci}%
  \BibitemOpen
  \bibfield  {author} {\bibinfo {author} {\bibfnamefont {O.}~\bibnamefont
  {Pavlosiuk}}\ and\ \bibinfo {author} {\bibfnamefont {D.}~\bibnamefont
  {Kaczorowski}},\ }\href {\doibase 10.1038/s41598-018-29545-w} {\bibfield
  {journal} {\bibinfo  {journal} {Scientific Reports}\ }\textbf {\bibinfo
  {volume} {8}},\ \bibinfo {pages} {11297} (\bibinfo {year}
  {2018})}\BibitemShut {NoStop}%
\bibitem [{\citenamefont {Pippard}(1990)}]{Pippard}%
  \BibitemOpen
  \bibfield  {author} {\bibinfo {author} {\bibfnamefont {A.}~\bibnamefont
  {Pippard}},\ }\href@noop {} {\emph {\bibinfo {title} {Magnetoresistance in
  Metals}}}\ (\bibinfo  {publisher} {Cambridge University Press},\ \bibinfo
  {address} {London},\ \bibinfo {year} {1990})\BibitemShut {NoStop}%
\bibitem [{\citenamefont {Singha}\ \emph {et~al.}(2017)\citenamefont {Singha},
  \citenamefont {Pariari}, \citenamefont {Satpati},\ and\ \citenamefont
  {Mandal}}]{singha2017}%
  \BibitemOpen
  \bibfield  {author} {\bibinfo {author} {\bibfnamefont {R.}~\bibnamefont
  {Singha}}, \bibinfo {author} {\bibfnamefont {A.~K.}\ \bibnamefont {Pariari}},
  \bibinfo {author} {\bibfnamefont {B.}~\bibnamefont {Satpati}}, \ and\
  \bibinfo {author} {\bibfnamefont {P.}~\bibnamefont {Mandal}},\ }\href
  {\doibase 10.1073/pnas.1618004114} {\bibfield  {journal} {\bibinfo  {journal}
  {Proceedings of the National Academy of Sciences of the United States of
  America}\ }\textbf {\bibinfo {volume} {114}},\ \bibinfo {pages} {2468}
  (\bibinfo {year} {2017})}\BibitemShut {NoStop}%
\bibitem [{\citenamefont {Ziman}(1962)}]{ziman}%
  \BibitemOpen
  \bibfield  {author} {\bibinfo {author} {\bibfnamefont {M.}~\bibnamefont
  {Ziman}},\ }\href@noop {} {\emph {\bibinfo {title} {Electrons and Phonons}}}\
  (\bibinfo  {publisher} {Clarendon Press},\ \bibinfo {address} {Oxford},\
  \bibinfo {year} {1962})\BibitemShut {NoStop}%
\bibitem [{\citenamefont {Burkov}(2017)}]{ChiralAprb}%
  \BibitemOpen
  \bibfield  {author} {\bibinfo {author} {\bibfnamefont {A.~A.}\ \bibnamefont
  {Burkov}},\ }\href {\doibase 10.1103/PhysRevB.96.041110} {\bibfield
  {journal} {\bibinfo  {journal} {Phys. Rev. B}\ }\textbf {\bibinfo {volume}
  {96}},\ \bibinfo {pages} {041110} (\bibinfo {year} {2017})}\BibitemShut
  {NoStop}%
\bibitem [{\citenamefont {Nandy}\ \emph {et~al.}(2017)\citenamefont {Nandy},
  \citenamefont {Sharma}, \citenamefont {Taraphder},\ and\ \citenamefont
  {Tewari}}]{ChiralAprl}%
  \BibitemOpen
  \bibfield  {author} {\bibinfo {author} {\bibfnamefont {S.}~\bibnamefont
  {Nandy}}, \bibinfo {author} {\bibfnamefont {G.}~\bibnamefont {Sharma}},
  \bibinfo {author} {\bibfnamefont {A.}~\bibnamefont {Taraphder}}, \ and\
  \bibinfo {author} {\bibfnamefont {S.}~\bibnamefont {Tewari}},\ }\href
  {\doibase 10.1103/PhysRevLett.119.176804} {\bibfield  {journal} {\bibinfo
  {journal} {Phys. Rev. Lett.}\ }\textbf {\bibinfo {volume} {119}},\ \bibinfo
  {pages} {176804} (\bibinfo {year} {2017})}\BibitemShut {NoStop}%
\bibitem [{\citenamefont {Wei}\ \emph {et~al.}(2023)\citenamefont {Wei},
  \citenamefont {Feng},\ and\ \citenamefont {Weng}}]{PHEsysm}%
  \BibitemOpen
  \bibfield  {author} {\bibinfo {author} {\bibfnamefont {Y.-W.}\ \bibnamefont
  {Wei}}, \bibinfo {author} {\bibfnamefont {J.}~\bibnamefont {Feng}}, \ and\
  \bibinfo {author} {\bibfnamefont {H.}~\bibnamefont {Weng}},\ }\href {\doibase
  10.1103/PhysRevB.107.075131} {\bibfield  {journal} {\bibinfo  {journal}
  {Phys. Rev. B}\ }\textbf {\bibinfo {volume} {107}},\ \bibinfo {pages}
  {075131} (\bibinfo {year} {2023})}\BibitemShut {NoStop}%
\bibitem [{\citenamefont {De~Ranieri}\ \emph {et~al.}(2008)\citenamefont
  {De~Ranieri}, \citenamefont {Rushforth}, \citenamefont {Výborný},
  \citenamefont {Rana}, \citenamefont {Ahmad}, \citenamefont {Campion},
  \citenamefont {Foxon}, \citenamefont {Gallagher}, \citenamefont {Irvine},
  \citenamefont {Wunderlich},\ and\ \citenamefont {Jungwirth}}]{AMR2008}%
  \BibitemOpen
  \bibfield  {author} {\bibinfo {author} {\bibfnamefont {E.}~\bibnamefont
  {De~Ranieri}}, \bibinfo {author} {\bibfnamefont {A.~W.}\ \bibnamefont
  {Rushforth}}, \bibinfo {author} {\bibfnamefont {K.}~\bibnamefont
  {Výborný}}, \bibinfo {author} {\bibfnamefont {U.}~\bibnamefont {Rana}},
  \bibinfo {author} {\bibfnamefont {E.}~\bibnamefont {Ahmad}}, \bibinfo
  {author} {\bibfnamefont {R.~P.}\ \bibnamefont {Campion}}, \bibinfo {author}
  {\bibfnamefont {C.~T.}\ \bibnamefont {Foxon}}, \bibinfo {author}
  {\bibfnamefont {B.~L.}\ \bibnamefont {Gallagher}}, \bibinfo {author}
  {\bibfnamefont {A.~C.}\ \bibnamefont {Irvine}}, \bibinfo {author}
  {\bibfnamefont {J.}~\bibnamefont {Wunderlich}}, \ and\ \bibinfo {author}
  {\bibfnamefont {T.}~\bibnamefont {Jungwirth}},\ }\href {\doibase
  10.1088/1367-2630/10/6/065003} {\bibfield  {journal} {\bibinfo  {journal}
  {New Journal of Physics}\ }\textbf {\bibinfo {volume} {10}},\ \bibinfo
  {pages} {065003} (\bibinfo {year} {2008})}\BibitemShut {NoStop}%
\bibitem [{\citenamefont {Yang}\ \emph {et~al.}(2020)\citenamefont {Yang},
  \citenamefont {Chang},\ and\ \citenamefont {Parkin}}]{PHEBiprr}%
  \BibitemOpen
  \bibfield  {author} {\bibinfo {author} {\bibfnamefont {S.-Y.}\ \bibnamefont
  {Yang}}, \bibinfo {author} {\bibfnamefont {K.}~\bibnamefont {Chang}}, \ and\
  \bibinfo {author} {\bibfnamefont {S.~S.~P.}\ \bibnamefont {Parkin}},\ }\href
  {\doibase 10.1103/PhysRevResearch.2.022029} {\bibfield  {journal} {\bibinfo
  {journal} {Phys. Rev. Res.}\ }\textbf {\bibinfo {volume} {2}},\ \bibinfo
  {pages} {022029} (\bibinfo {year} {2020})}\BibitemShut {NoStop}%
\bibitem [{\citenamefont {Yamada}\ and\ \citenamefont
  {Fuseya}(2021)}]{PHEBiprb}%
  \BibitemOpen
  \bibfield  {author} {\bibinfo {author} {\bibfnamefont {A.}~\bibnamefont
  {Yamada}}\ and\ \bibinfo {author} {\bibfnamefont {Y.}~\bibnamefont
  {Fuseya}},\ }\href {\doibase 10.1103/PhysRevB.103.125148} {\bibfield
  {journal} {\bibinfo  {journal} {Phys. Rev. B}\ }\textbf {\bibinfo {volume}
  {103}},\ \bibinfo {pages} {125148} (\bibinfo {year} {2021})}\BibitemShut
  {NoStop}%
\bibitem [{\citenamefont {Liu}\ \emph {et~al.}(2024{\natexlab{b}})\citenamefont
  {Liu}, \citenamefont {Zhang}, \citenamefont {Fang}, \citenamefont {Weng},\
  and\ \citenamefont {Wu}}]{liu2024}%
  \BibitemOpen
  \bibfield  {author} {\bibinfo {author} {\bibfnamefont {Z.}~\bibnamefont
  {Liu}}, \bibinfo {author} {\bibfnamefont {S.}~\bibnamefont {Zhang}}, \bibinfo
  {author} {\bibfnamefont {Z.}~\bibnamefont {Fang}}, \bibinfo {author}
  {\bibfnamefont {H.}~\bibnamefont {Weng}}, \ and\ \bibinfo {author}
  {\bibfnamefont {Q.}~\bibnamefont {Wu}},\ }\href
  {https://arxiv.org/abs/2401.15146} {\enquote {\bibinfo {title}
  {First-principles methodology for studying magnetotransport in magnetic
  materials},}\ } (\bibinfo {year} {2024}{\natexlab{b}}),\ \Eprint
  {http://arxiv.org/abs/2401.15146} {arXiv:2401.15146 [cond-mat.mtrl-sci]}
  \BibitemShut {NoStop}%
\bibitem [{\citenamefont {Zeng}\ \emph {et~al.}(2006)\citenamefont {Zeng},
  \citenamefont {Yao}, \citenamefont {Niu},\ and\ \citenamefont
  {Weitering}}]{zeng2006linear}%
  \BibitemOpen
  \bibfield  {author} {\bibinfo {author} {\bibfnamefont {C.}~\bibnamefont
  {Zeng}}, \bibinfo {author} {\bibfnamefont {Y.}~\bibnamefont {Yao}}, \bibinfo
  {author} {\bibfnamefont {Q.}~\bibnamefont {Niu}}, \ and\ \bibinfo {author}
  {\bibfnamefont {H.~H.}\ \bibnamefont {Weitering}},\ }\href
  {https://doi.org/10.1103/PhysRevLett.96.037204} {\bibfield  {journal}
  {\bibinfo  {journal} {Physical review letters}\ }\textbf {\bibinfo {volume}
  {96}},\ \bibinfo {pages} {037204} (\bibinfo {year} {2006})}\BibitemShut
  {NoStop}%
\bibitem [{\citenamefont {Zhao}\ \emph {et~al.}(2023)\citenamefont {Zhao},
  \citenamefont {Jiang}, \citenamefont {Yang}, \citenamefont {Wang},
  \citenamefont {Shi}, \citenamefont {Tian}, \citenamefont {Li}, \citenamefont
  {Liu},\ and\ \citenamefont {Wu}}]{zhao2023magnetotransport}%
  \BibitemOpen
  \bibfield  {author} {\bibinfo {author} {\bibfnamefont {J.}~\bibnamefont
  {Zhao}}, \bibinfo {author} {\bibfnamefont {B.}~\bibnamefont {Jiang}},
  \bibinfo {author} {\bibfnamefont {J.}~\bibnamefont {Yang}}, \bibinfo {author}
  {\bibfnamefont {L.}~\bibnamefont {Wang}}, \bibinfo {author} {\bibfnamefont
  {H.}~\bibnamefont {Shi}}, \bibinfo {author} {\bibfnamefont {G.}~\bibnamefont
  {Tian}}, \bibinfo {author} {\bibfnamefont {Z.}~\bibnamefont {Li}}, \bibinfo
  {author} {\bibfnamefont {E.}~\bibnamefont {Liu}}, \ and\ \bibinfo {author}
  {\bibfnamefont {X.}~\bibnamefont {Wu}},\ }\href
  {https://doi.org/10.1103/PhysRevB.107.L060408} {\bibfield  {journal}
  {\bibinfo  {journal} {Physical Review B}\ }\textbf {\bibinfo {volume}
  {107}},\ \bibinfo {pages} {L060408} (\bibinfo {year} {2023})}\BibitemShut
  {NoStop}%
\end{thebibliography}%
\bibliographystyle{apsrev4-1}      

\end{document}